\documentclass[journal]{IEEEtran}
\usepackage[T1]{fontenc}
\usepackage[latin9]{inputenc}
\usepackage{color}
\usepackage{array}
\usepackage{float}
\usepackage{booktabs}
\usepackage{textcomp}
\usepackage{multirow}
\usepackage{amsmath}
\usepackage{amsthm}
\usepackage{amssymb}
\usepackage{graphicx}
\usepackage[unicode=true,
 bookmarks=true,bookmarksnumbered=true,bookmarksopen=true,bookmarksopenlevel=1,
 breaklinks=false,pdfborder={0 0 0},pdfborderstyle={},backref=false,colorlinks=false]
 {hyperref}
\hypersetup{pdftitle={Your Title},
 pdfauthor={Your Name},
 pdfpagelayout=OneColumn, pdfnewwindow=true, pdfstartview=XYZ, plainpages=false}

\makeatletter

\providecommand{\tabularnewline}{\\}
\floatstyle{ruled}
\newfloat{algorithm}{tbp}{loa}
\providecommand{\algorithmname}{Algorithm}
\floatname{algorithm}{\protect\algorithmname}

\let\oldforeign@language\foreign@language
\DeclareRobustCommand{\foreign@language}[1]{%
\lowercase{\oldforeign@language{#1}}}
\theoremstyle{remark}
\newtheorem{rem}{\protect\remarkname}
\theoremstyle{definition}
\newtheorem{defn}{\protect\definitionname}
\theoremstyle{plain}
\newtheorem{lem}{\protect\lemmaname}
\theoremstyle{plain}
\newtheorem{thm}{\protect\theoremname}

\usepackage[caption=false,font=footnotesize]{subfig}
\usepackage{textgreek}
\usepackage{algorithm,algpseudocode}
\theoremstyle{definition}
\newtheorem{assumption}{Assumption}
\newtheorem{condition}{Condition}

\@ifundefined{showcaptionsetup}{}{%
 \PassOptionsToPackage{caption=false}{subfig}}
\usepackage{subfig}
\makeatother

\providecommand{\definitionname}{Definition}
\providecommand{\lemmaname}{Lemma}
\providecommand{\remarkname}{Remark}
\providecommand{\theoremname}{Theorem}

\begin{document}
\bstctlcite{BSTcontrol}
\title{Reinforcement Learning Control of Robotic Knee with Human in the Loop
by Flexible Policy Iteration}
\author{Xiang~Gao, Jennie~Si,~\IEEEmembership{Fellow,~IEEE,} Yue~Wen,
Minhan~Li, and~He (Helen)~Huang,~\IEEEmembership{Senior~Member,~IEEE}\thanks{This work was partly supported by National Science Foundation \#1563454,
\#1563921, \#1808752 and \#1808898.}\thanks{X.~Gao and J.~Si are with the the Department of Electrical, Computer,
and Energy Engineering, Arizona State University, Tempe, AZ 85287
USA (e-mail: \protect\href{mailto:xgao29@asu.edu}{xgao29@asu.edu};
\protect\href{mailto:si@asu.edu}{si@asu.edu} (correspondence)).}\thanks{Y.~Wen, M.~Li and H.~Huang are with the Department of Biomedical
Engineering, North Carolina State University, Raleigh, NC 27695 USA,
and also with the University of North Carolina at Chapel Hill, Chapel
Hill, NC 27599 USA (e-mail: \protect\href{mailto:ywen3@ncsu.edu}{ywen3@ncsu.edu};
\protect\href{mailto:mli37@ncsu.edu}{mli37@ncsu.edu}; \protect\href{mailto:hhuang11@ncsu.edu}{hhuang11@ncsu.edu}).}}
\markboth{}{Xiang Gao \MakeLowercase{\emph{et al.}}: Reinforcement Learning
Control of Robotic Knee with Human in the Loop by Flexible Policy
Iteration}
\maketitle
\begin{abstract}
We are motivated by the real challenges presented in a human-robot
system to develop new designs that are efficient at data level and
with performance guarantees such as stability and optimality at systems
level. Existing approximate/adaptive dynamic programming (ADP) results
that consider system performance theoretically are not readily providing
practically useful learning control algorithms for this problem; and
reinforcement learning (RL) algorithms that address the issue of data
efficiency usually do not have performance guarantees for the controlled
system. This study fills these important voids by introducing innovative
features to the policy iteration algorithm. We introduce flexible
policy iteration (FPI), which can flexibly and organically integrate
experience replay and supplemental values from prior experience into
the RL controller. We show system level performances including convergence
of the approximate value function, (sub)optimality of the solution,
and stability of the system. We demonstrate the effectiveness of the
FPI via realistic simulations of the human-robot system. It is noted
that the problem we face in this study may be difficult to address
by design methods based on classical control theory as it is nearly
impossible to obtain a customized mathematical model of a human-robot
system either online or offline. The results we have obtained also
indicate the great potential of RL control to solving realistic and
challenging problems with high dimensional control inputs. 
\end{abstract}


\IEEEpeerreviewmaketitle{}

\section{Introduction}

\IEEEPARstart{R}{obotic} knees are wearable robots that assist individuals
with lower limb amputation to regain the ability of walking \cite{Sup2009a,Rouse2014Clutch}.
This type of robotic prosthesis relies on mimicking how biological
joints generate torques to enable the robotic knee motion for an amputee
user. The device is programmed to adjust the knee joint impedance
values according to the mechanical sensors in the prosthesis. The
intrinsic controllers embedded in the devices provide baseline automatic
control of joint torques. As human users differ from weight to size
and are of different life-style needs, extrinsic control in the form
of providing impedance parameter settings are required to customize
the device to meet individual user\textquoteright s physical and life
style needs. 

\noindent \emph{Related Approaches.} While such new powered device
signifies the future of rehabilitation, and it has brought excitement
into the biomechanics fields, fitting the device to a human user automatically
remains a major challenge to unleash the full potential of the robotic
device. Few technologies are currently available. The only functioning
solution relies on multiple sessions of manually tuning a small subset
of the impedance parameters one at a time, unable to account for the
interacting effects of the parameters during each tuning session.
This highly heuristic approach is time consuming, costly, and not
scalable to reaching the full potential of this powerful robotic device.

Some research has gone into reducing the intensity of labor for parameter
tuning. One such idea directly reduces the number of configurable
parameters \cite{Simon2014,Gregg2014}. In return, significant domain
knowledge and tuning time are still required, and it is not clear
if such an approach will remain effective for each unique individual
of the amputee population. In \cite{Huang2016}, an expert system
was developed to hard code the prosthetists\textquoteright{} tuning
decisions into configuring the control parameters. This open-loop
control approach is not expected to scale well to more joints or to
more tasks. Other approaches include estimating the control impedance
parameters with either a musculoskeletal model \cite{Pfeifer2012}
or measurements of biological joint impedance \cite{Rouse2014,Tucker2017}.
These methods have not been validated and it is questionable if they
are feasible as the biomechanics and the joint activities of amputees
are fundamentally different from those of the able-bodied population.
A recent continuous tracking approach was proposed based on extremum
seeking, aka a convex optimization solution, for seeking impedance
parameters \cite{Kumar2020}. The idea applied as a concept to knee
and ankle joints by automatically tuning the proportion gain of a
PD controller. It is yet to see direct results of leg motion performance
either in simulation or by human testing.

As those methods all have their fundamental limitations in principled
ways, new approaches of configuring the prosthesis control parameters
are needed. Even though controlling exoskeleton is not quite the same
problem as configuring prosthesis control parameters because of the
fundamental and physiological differences between able-bodied and
amputee populations, it is still worth mentioning that several optimization
techniques have been proposed for the former. In \cite{Koller-RSS-16},
the authors used gradient descent method to determine an optimal control
onset time of an ankle exoskeleton device. Their goal was to minimize
the metabolic effort from respiratory measurements and thus improve
gait efficiency. The authors of \cite{Zhang2017HumanintheloopOO}
also studied ankle exoskeleton towards minimizing the metabolic effort,
where in the study, they used an evolution strategy to optimize four
control parameters in the ankle device. Ding et al. \cite{Ding2018}
applied Bayesian optimization to place two control parameters of hip
extension assistance. While this idea and those discussed are interesting
for exoskeleton, they may not extend to robotic prosthesis parameter
tuning problem as these algorithms have only been demonstrated for
less or equal to a 5-dimensional control parameter space. In addition
to scalability, they have not shown feasibility in addressing challenges
such as adapting to user\textquoteright s changing physical condition
or change in use environments. Additionally, the metabolic cost objective
used in control design of exoskeletons for able bodied individuals
is unlikely useful for amputees using a prosthetic device.

\medskip{}

\noindent \emph{Problem Challenges.} Several unique characteristics
of the human-robot system are responsible for the challenges we face
when configuring the robotic devices. First, fundamental principles
and mechanisms of how the human and prosthesis interact still is not
known. Therefore, it is not feasible to apply control design approaches
that rely on a mathematical description of the human-prosthesis dynamics.
Second, lower limb prosthesis tuning is commonly implemented in a
finite state machine impedance control (FS-IC) framework \cite{Hogan1985}
because studies have suggested that humans actually control the joint
impedance of the leg when walking \cite{Geyer2003,Shamaei2013}. The
FS-IC involves multiple configurable parameters or control inputs,
from 12 to 15 for knee prosthesis for level ground walking \cite{Sup2009a,Rouse2014Clutch,Liu2014a},
and the number of parameters grows rapidly for increased number of
joints and locomotion modes up to multiple dozens. Third, the impedance
control design has to ensure the safety and stability of the human
user at all times. Additionally, because of a human user is in the
loop during the tuning process, it is highly desirable for the control
design approach to be data and time efficient to reduce the discomfort
caused by adjusting the control parameters. Addressing these challenges
simultaneously requires us to look beyond classical control systems
theory and control systems engineering as well as the state-of-the-art
robotics science and engineering that have been successful at controlling
mechanical robots.

\medskip{}

\noindent \emph{Reinforcement Learning and Adaptive Dynamic Programming.}
The reinforcement learning (RL) based adaptive optimal control is
naturally appealing to solve the above described challenges. As is
well known, deep RL, including several policy search methods and Deep
Q-Network (DQN), have shown unprecedented successes in solving difficult,
sequential decision-making problems, such as those in robotics applications
\cite{Levine2016}, Atari games \cite{Mnih2015}, the game of Go \cite{Silver2016a,Silver2017b}
and energy efficient data center \cite{Farahnakian2014}. Yet, a few
challenges remain if these results are to be extended to situations
where there is no abundance of data, the problems involve continuous
state and control variables, uncertainties as well as sensor and actuator
noise are inevitable, and system performances such as optimality and
stability have to be satisfied. RL based adaptive optimal control
approaches, or adaptive/approximate dynamic programming (ADP) \cite{Si2004a,Frank2012},
is a promising alternative as they have demonstrated their capability
of learning from data measurements in an online or offline manner
in several realistic application problems including large-scale control
problems, such as power system stability enhancement \cite{Lu2008}-\nocite{Lu2008,Guo2015,Guo2016a}\cite{Guo2016a},
and Apache helicopter control \cite{Enns2002,Enns2003}. Note however,
those problems do not face the data and time efficiency challenge.

At the heart of the ADP methods is the idea of providing approximating
solutions to the Bellman equation of optimal control problems. In
our previous work \cite{Wen2016a}\nocite{Wen2016a,Wen2017,Wen2019}-\cite{Wen2019},
we demonstrated the feasibility of ADP, specifically direct heuristic
dynamic programming (dHDP) \cite{Si2001}, for personalizing robotic
knee control, to address similar challenges we face here. The dHDP
is an online RL algorithm based on stochastic gradient descent, which
in its generic form, is not optimized for fast learning \cite{Lagoudakis2004}.
It is also worth mentioning that, the generic dHDP without imposing
further conditions \cite{Liu2012a} have not shown its control law
to be stable during learning. It is therefore necessary to take these
limitations into design considerations especially for the current
application. 

\medskip{}

\noindent \emph{Policy Iteration RL Control.} AlphaGo Zero \cite{Silver2017b}
is a policy iteration-based reinforcement learning algorithm. It started
tabula rasa, achieved superhuman performance after only a few days
of training. This result is inspiring. Yet, how to make a classic
policy iteration algorithm applicable to solving controls problem
that requires data and time efficiency as well as system stability
and optimal performance remains a challenge. ADP is a promising adaptive
optimal control framework to address general nonlinear control problems.
But few results are available to demonstrate successful applications
to real controls problems while meeting the data and time efficiency
requirements. Some motivating and important theoretical works are
as follows, but they do not directly involve data-level design approaches
to solving realistic and complex problems. In \cite{WGao2018}\nocite{WGao2018,YJiang2015,bian2014}-\cite{bian2014},
the authors examined continuous time systems of different constraining
nonlinearity forms (affine systems, for example) under respective
state and control conditions. For discrete-time systems, \cite{JiangLewis2019}
deals with a class of linear systems, \cite{Al-Tamimi2008} deals
with an affine nonlinear system with learning convergence proof, \cite{Mu2017}
deals with nonlinear systems with learning convergence proof and optimality
performance, However, system stability was not provided in \cite{Al-Tamimi2008,Mu2017}.
Stable iterative control policies of PI have been studied in \cite{Liu2014f}-\nocite{Liu2014f,Wei2015b,Wei2016n,Guo2017}\cite{Guo2017}
for discrete-time nonlinear systems of different forms. However, only
within the generic PI formulation without any consideration of integrating
prior knowledge. Clearly, we need a practical, data level design algorithm
that is useful for applications while retaining important, system
level performance properties such as optimality and stability.

\medskip{}

\noindent \emph{Using Prior Information.} Two design ideas to account
for data and time efficiency are intuitively useful \textendash{}
experience replay and value function shaping. We innovatively develop
both ideas and organically integrate them into our FPI algorithms.

Experience replay (ER) \cite{Lin1992} is a practically effective
approach to improving sample efficiency for off-policy RL methods.
In ER, past experiences (samples) generated under different behavior
policies are stored in a memory buffer and selected repeatedly for
evaluating the approximated value function. Even though ER has been
adopted and analyzed extensively, shown below, current results do
not simultaneously fulfill our design requirement of data level efficiency
and performance guarantee simultaneously. Empirical studies have demonstrated
successes of several ER algorithms. Selective experience replay (SER)
\cite{Isele2018}, prioritized experience replay (PER) \cite{Schaul2015,Horgan2018}
and hindsight experience replay (HER) \cite{Andrychowicz2017} have
shown their capability in improving sample efficiency in deep RL.
Specifically, SER strategically selects which experiences will be
stored so that the distribution on the memory buffer can match the
global distribution in the training dataset. In contrast, PER selects
which experiences to be replayed in the memory buffer. That is to
say, the important samples will be replayed more frequently. For multi-goal
RL, HER reutilizes past failure experiences which may benefit for
another goal, so that the overall multi-goal performance can be improved.The
ER idea has also been considered in ADP in different capacities \cite{Adam2012a}-\nocite{Luo2018b,chowdhary2013concurrent,Modares2014,vamvoudakis2016asymptotically,Zhao2016,Yang2019}\cite{Yang2019}.
Beyond the results in \cite{Lin1992}-\cite{Andrychowicz2017}, the
works in \cite{Adam2012a} and \cite{Luo2018b} also empirically demonstrated
that the simple experience replay idea, without prioritization or
other means of selecting samples, can be implemented with Q-learning
and actor-critic ADP algorithms to improve sample efficiency. Some
analytical studies \cite{chowdhary2013concurrent}-\cite{Yang2019}
about ER have carried out for specific systems and under specific
conditions, which are not sufficiently general to be applied to our
human-robot system. Specifically, ER, or concurrent learning, was
proposed to replace the persistence of excitation (PE) condition for
uncertain linear dynamical systems \cite{chowdhary2013concurrent},
partially unknown constrained input systems \cite{Modares2014}, known
deterministic nonlinear systems \cite{vamvoudakis2016asymptotically},
non-zero sum games based on model identifiers of the game systems
\cite{Zhao2016} and decentralized event-triggered control of interconnected
systems in affine form with uncertain interconnections \cite{Yang2019}.
Apparently, existing results are not readily extended to address the
challenges just discussed in the above.

The idea of using prior experience to bootstrap learning is intuitive
as it intends to capture knowledge from a related task to help save
data and learning time. Research has shown significant improvement
in data and time efficiency by using prior experience as compared
to learning from scratch \cite{Taylor2009a,abel2018policy,Lazaric2012}.
At present, most approaches rely on identifying and applying a good
initial policy or initial value function to \textquotedblleft guide\textquotedblright{}
learning in order to reduce the policy search space as a means of
reducing data complexity and saving learning time \cite{Taylor2005,Fernandez2006,Griffith2013,Konidaris2006}.
A handful of results attempted formal treatment of utilizing prior
knowledge to boost learning, yet they are only for special considerations.
\cite{Ng1999} is an early work that considered boosting the value
function or shaping the reward function. But the focus was to ensure
policy invariance, not from a perspective of saving data and training
time. \cite{Mann2012} provided a theoretical sample complexity framework,
yet it is unclear how the results could directly impact the development
of practically useful algorithms. \cite{abel2018policy} proved a
probably approximately correct (PAC) guarantee for a scheme that uses
an optimistic value function initialization, and the authors only
demonstrated their approach using simple examples. 

\medskip{}

\noindent \emph{Contributions and Significance.} From the above discussions,
it is clear that practical, data level design algorithms with important,
system level performance properties are still lacking. In this study,
we introduce flexible policy iteration (FPI) to address the challenges.
The flexibility of our proposed FPI is within the following three
aspects. First, the way it collects and uses data, i.e., data preparation
(Table I), to permit learning from samples generated from either current
behavior policies or different policies within an experience replay
framework. Second, the way it deals with prior knowledge is flexible
as it allows learning from prior knowledge in the form of a supplemental
value based on previous data collection experiments. Third, the implementation
of FPI is flexible as the approximate value function can be obtained
by a conventional least-square solution or by a weighted least-square
solution with or without prioritized samples. With such a flexible
framework, a designer of an adaptive optimal controller can customize
his/her algorithmic approach to fit specific applications needs. 

In summary, we are motivated by the real challenges presented in a
human-robot system to develop new designs that are efficient at data
level and with performance guarantees at systems level. Successful
applications of policy iteration, such as AlphaGo Zero, are inspiring
but did not account for either data efficiency or system stability.
Existing ADP results that consider system performance theoretically
are not readily providing practically useful learning control algorithms.
This study fills these important voids. Specifically, our contributions
are as follows. 
\begin{itemize}
\item First, based on the classic policy iteration framework, we introduce
several flexibilities at data level, each or all of which can be customized
to meet the design and problem solving needs. Our innovative development
of experience replay and supplemental values, and organic integration
of them into the policy evaluation, have provided practically useful
design tools. 
\item Second, we not only introduce FPI as an iterative learning procedure,
but we also provide qualitative analysis of FPI for its stabilizing
control laws, convergence of the value function and achieving Bellman
optimality approximately. 
\item Third, we provide extensive simulation studies to validate what we
intended for the FPI: to automatically configure 12 impedance parameters
as prosthesis control inputs to enable continuous walking of amputee
subjects. 
\end{itemize}
It is noted that, what we have obtained in this study would have been
difficult or impossible for designs based on classic control theory.
Our results reported here represent the state of the art in automatic
configuration of powered prosthetic knee devices. This result may
potentially lead to practical use of the FPI in clinics. In turn,
this can significantly reduce health care cost and improve the quality
of life for the transfemoral amputee population in the world. 

The remaining of this paper is organized as follows. Section II describes
the human-prosthesis system and formulates the human-prosthesis tuning/configuration
problem. Section III presents the FPI framework for online control
of prosthetic knee. Section IV analyzes the converging properties
of FPI. Section V presents the experimental results of the FPI-tuner
under different configurations, and a comprehensive comparison to
other existing RL methods. Discussions and conclusion are presented
in Section VI.

\section{Human-Robot System\label{sec:Human-Prosthesis-Simulation-Plat}}

\begin{figure*}[t]
\begin{centering}
\subfloat[]{\begin{centering}
\includegraphics[width=1.2\columnwidth]{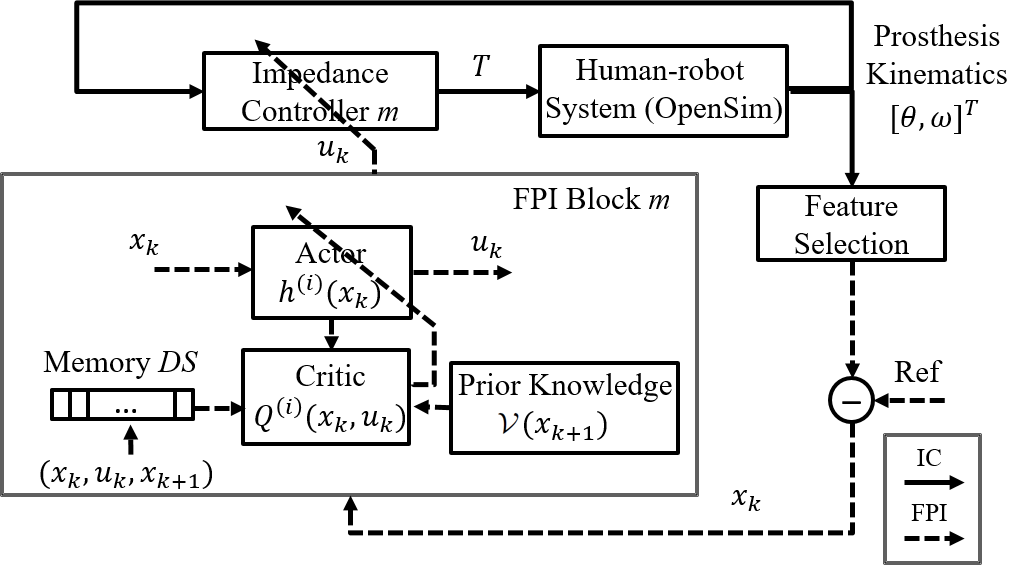}
\par\end{centering}

}\subfloat[]{\includegraphics[width=0.8\columnwidth]{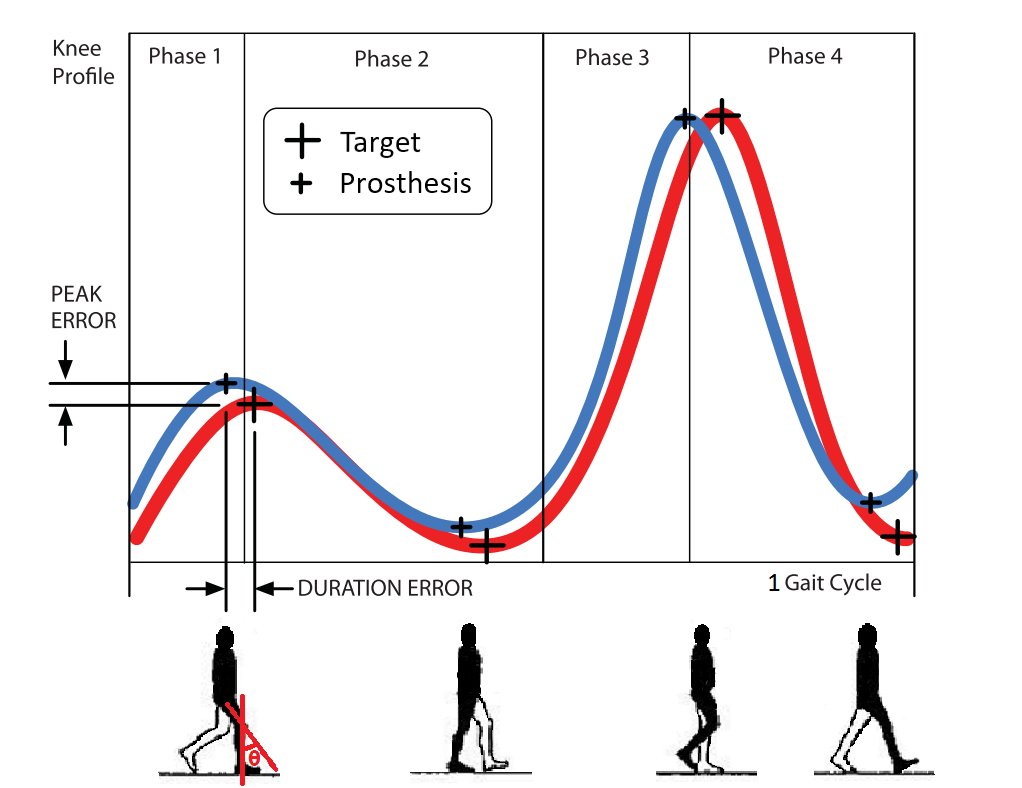}

}
\par\end{centering}
\caption{{(a) Block diagram of the human-robot system with
a RL controlled robotic knee. The impedance control loop (IC loop,
top) generates torque $T$ according to \eqref{eq:torque}. The FPI-based
parameter update loop (FPI loop) adjusts impedance control parameters
for each phase $m$ after every gait cycle $k$. Four identical RL
blocks ($m=1,2,3,4$) are needed for the four IC control phases. (b)
Illustration of the four phases of a gait cycle: the differences between
the target and prosthesis knee profiles form the states peak error
and duration error of the four respective phases.} \label{fig:The-block-diagram}}
\end{figure*}

In this study, RL is applied to automatically adjust impedance parameter
settings within a finite state impedance control (FS-IC) framework,
where a gait cycle is divided into four phases to represent different
modes of stance and swing \cite{Sup2009a,Rouse2014,Liu2014a} (Fig.
\ref{fig:The-block-diagram}(b)): stance flexion phase (STF, $m=1$),
stance extension phase (STE, $m=2$), swing flexion (SWF, $m=3$)
and swing extension (SWE, $m=4$). Transitions between phases are
triggered by the ground reaction force (GRF), knee joint angle, and
knee joint angular velocity measured from the prosthesis. Variance
in a certain phase will affect the subsequent phases \cite{Wen2017b}.
Fig. \ref{fig:The-block-diagram}(a) shows an FS-IC based human-prosthesis
system and how our proposed reinforcement learning control is integrated
into the system. There are two control loops running at different
frequencies. The intrinsic, impedance control (IC) loop generates
knee joint torque $T$ at 300 Hz following the impedance control law
\eqref{eq:torque}, given an impedance control parameter setting defined
in (1). For each of the gait phases $m=1,2,3,4,$ there is a respective
FPI controller. In each gait cycle $k$ for each phase $m$, state
$x_{k}$ is formed using peak knee angle $P_{k}$ and gait phase duration
$D_{k}$ measures as shown in Fig. \ref{fig:The-block-diagram}(b).

\subsection{The Intrinsic Impedance Control Loop \label{subsec:Impedance-Control-Loop}}

Refer to the top IC control loop of Fig. 1(a). During gait cycle $k$,
for each FS-IC control phase $m$ $(m=1,2,3,4)$, robotic knee control
requires three control parameters, namely stiffness $K_{m,k}$, damping
coefficient $B_{m,k}$ and equilibrium position $(\theta_{e})_{m,k}$.
In vector form, the control parameter settings are represented as 

\begin{equation}
I_{m,k}=[K_{m,k},B_{m,k},(\theta_{e})_{m,k}]^{T}\in\mathbb{R}^{3}.\label{eq:imp}
\end{equation}
The prosthetic knee motor generates a knee joint torque $T\in\mathbb{R}$
from the knee joint angle $\theta$ and angular velocity $\omega$
according to the following impedance control law
\begin{equation}
T_{k}=K_{k}(\theta-(\theta_{e})_{k})+B_{k}\omega.\label{eq:torque}
\end{equation}
Without loss of generality, we drop the subscript $m$ in the rest
of the paper because all four impedance controllers and their respective
FPI blocks share the same structure, although RL controller for each
phase has its own coefficients to generate impedance control settings
(1). The FPI controller then updates the IC parameter settings \eqref{eq:imp}
for the next gait cycle $k+1$ as 
\begin{equation}
I_{k+1}=I_{k}+u_{k},\label{eq:imp-update}
\end{equation}
where $u_{k}\in\mathbb{R}^{3}$ is the control input generated from
the FPI block. 

\subsection{Impedance Parameter Update Loop by FPI\label{subsec:Parameter-Update-Loop}}

Refer to the bottom control loop of Fig. 1(b). For each phase $m$
during gait cycle $k$, the $m$th FPI controllers is enabled to update
the IC parameters. After each gait cycle $k$, the peak knee angle
$P_{k}\in\mathbb{R}$ and phase duration $D_{k}\in\mathbb{R}$ are
selected by the feature selection module. Specifically, peak knee
angle $P_{k}$ is the maximum or minimum knee angle in each phase,
and phase duration $D_{k}$ is the time interval between two consecutive
peaks (Fig. \ref{fig:The-block-diagram}(b)). A reference trajectory
of the knee joint that resembles a normal human walking pattern \cite{Sup2009a,Martinez-villalpando2009}
is used in this study. Subsequently we can also determine target peak
angle $P_{k}^{'}\in\mathbb{R}$ and phase duration $D_{k}^{'}\in\mathbb{R}$.
For each RL controller in a respective phase, its state variable $x_{k}$
is defined using peak error $\Delta P_{k}\in\mathbb{R}$ and duration
error $\Delta D_{k}\in\mathbb{R}$ as 
\begin{equation}
x_{k}=[\text{\ensuremath{\Delta}}P_{k},\text{\ensuremath{\Delta}}D_{k}]{}^{T}=[P_{k}-P_{k}^{'},D_{k}-D_{k}^{'}]{}^{T},\label{eq:ADP-state}
\end{equation}
and its control $u_{k}$ consists of increments to the IC parameters,
\begin{equation}
u_{k}=[\Delta K_{k},\Delta B_{k},(\text{\ensuremath{\Delta}}\theta_{e}\text{)}{}_{k}]^{T}.\label{eq:u_k}
\end{equation}

\section{Flexible Policy Iteration\label{sec:Batch-Approximate-Policy}}

Consider the human-robot, i.e., the amputee-prosthesis, system as
a discrete time nonlinear system with unknown dynamics, 
\begin{equation}
x_{k+1}=F(x_{k},u_{k}),\hphantom{}k=0,1,2,\text{\dots}\label{eq:system}
\end{equation}
where action $u_{k}$ of the form described in \eqref{eq:u_k} is
determined according to policy $h$ as 
\begin{equation}
u_{k}=h(x_{k}).\label{eq:uk}
\end{equation}
In \eqref{eq:system}, the domain of $F(x_{k},u_{k})$ is denoted
as $\mathcal{D}\triangleq\{(x,u)|x\in\mathfrak{\mathcal{X}},u\in\mathcal{U}\}$,
where $\mathcal{X}$ and $\mathcal{U}$ are compact sets with dimensions
of $N_{x}$ and $N_{u}$, respectively. In the human-robot system
under consideration, $F$ represents the kinematics of the robotic
knee, which is affected by both the human wearer and also the RL controller.
Because of a human in-the-loop, an explicit mathematical model as
\eqref{eq:system} is intractable or impossible to obtain. 

\subsection{The Policy Iteration Framework}

To assist our development of the proposed flexible policy iteration
(FPI), we summarize the notation and the basic framework of a standard
policy iteration algorithm for discrete time systems next.

The RL control design objective is to derive an optimal control law
via learning from observed data along the human-robot system dynamics.
Consider a control policy $h(x_{k})$, we define the state-action
Q-value function as

\begin{equation}
\check{Q}(x_{k},u_{k})=\check{U}(x_{k},u_{k})+\sum_{j=1}^{\infty}\check{U}(x_{k+j},h(x_{k+j}))\label{eq:Qvalue}
\end{equation}
where $\check{U}(x_{k},u_{k})$ is the stage cost or instantaneous
cost function. Note that the $\check{Q}(x_{k},u_{k})$ value is a
performance measure when action $u_{k}$ is applied at state $x_{k}$
and the control policy $h$ is followed thereafter. The form of $\check{Q}(x_{k},u_{k})$
in \eqref{eq:Qvalue} implies that we formulate the optimal control
problem of robotic knee as a discrete time, infinite horizon and undiscounted
optimization problem. Our solution framework is data-driven, not model-based. 

Our design approach requires the following assumption.\theoremstyle{definition}\begin{assumption}The
system $F(x_{k},u_{k})$ \eqref{eq:system} is controllable; the system
state $x_{k}=0$ is an equilibrium state of system \eqref{eq:system}
under the control $u_{k}=0$, i.e., $F(0,0)=0$; the feedback control
$u_{k}=h(x_{k})$ satisfies $u_{k}=h(x_{k})=0$ for $x_{k}=0$; the
stage cost function $\check{U}(x_{k},u_{k})$ in $x_{k}$ and $u_{k}$
is positive definite.\end{assumption}Assumption 1 is satisfied in
the robotic knee control problem due to our construction of the system
states and RL control \eqref{eq:imp-update} based on the biomechanics
of human locomotion.

The Q-value function in \eqref{eq:Qvalue} satisfies the following
Bellman equation,
\begin{equation}
\check{Q}(x_{k},u_{k})=\check{U}(x_{k},u_{k})+\check{Q}(x_{k+1},h(x_{k+1})).\label{eq:Bellman equation for Q-value}
\end{equation}
An optimal control is the one that stabilizes the system in \eqref{eq:system}
while minimizing the value function \eqref{eq:Qvalue} according to
Bellman optimality. The optimal value function is therefore of the
form
\begin{equation}
\begin{aligned}Q^{*}(x_{k},u_{k})=\check{U}(x_{k},u_{k})+\underset{u_{k+1}}{\min}Q^{*}(x_{k+1},u_{k+1})\end{aligned}
\label{eq:Bellman optimality}
\end{equation}
or

\begin{equation}
h^{*}(x_{k})=\underset{u_{k}}{\arg\min}Q^{*}(x_{k},u_{k}),\label{eq:optimal_policy}
\end{equation}
\begin{equation}
Q^{*}(x_{k},u_{k})=\check{U}(x_{k},u_{k})+Q^{*}(x_{k+1},h^{*}(x_{k+1})),\label{eq:Bellman optimality 2}
\end{equation}
where $h^{*}(x_{k})$ denotes the optimal control policy.

Consider an iterative value function $\check{Q}^{(i)}(x_{k},u_{k})$
and a control policy $\check{h}^{(i)}(x_{k})$, the policy iteration
algorithm proceeds by iterating the follow two steps for $i=0,1,2,\ldots$:

\subsubsection*{Policy Evaluation}

\begin{align}
\check{Q}^{(i)}(x_{k},u_{k})= & \check{U}(x_{k},u_{k})+\check{Q}^{(i)}(x_{k+1},\check{h}^{(i)}(x_{k+1})).\label{eq:policy-evaluation-original}
\end{align}
The above policy evaluation step \eqref{eq:policy-evaluation-original}
is based on the Bellman equation \eqref{eq:Bellman equation for Q-value}.

\subsubsection*{Policy Improvement}

\begin{equation}
\check{h}^{(i+1)}(x_{k})=\underset{u_{k}}{\arg\min}\check{Q}^{(i)}(x_{k},u_{k}).\label{eq:policy-improvement-original}
\end{equation}

Motivated by the favorable properties of policy iteration in MDP problems,
such as monotonically decreasing value, and demonstrated feasibility
in solving realistic engineering problems \cite{Guo2015,Guo2016a},
we further develop the policy evaluation step to achieve data efficiency,
easy implementation, and importantly, effectively solving realistic
and complex problems. 

\subsection{Flexible Policy Iteration with Supplemental Value}

We first consider a flexible use of prior information, which we expect
to improve learning efficiency in data and time. Our approach entails
a supplemental value $\mathcal{V}(x_{k})$ which can be obtained from
an FPI solution based on past experience such as a robotic knee control
experiment involving a similar subject(s) previously. For $i=0,1,2,\ldots$
we define a new augmented cost-to-go $Q^{(i)}(x_{k},u_{k})$ based
on a supplemental value $\mathcal{V}(x_{k})$,

\begin{equation}
\begin{aligned}Q^{(i)}(x_{k},u_{k})= & \check{U}(x_{k},u_{k})+\sum_{j=1}^{\infty}\check{U}(x_{k+j},h^{(i)}(x_{k+j}))\\
 & +\sum_{j=0}^{\infty}\alpha_{i}\mathcal{V}(x_{k+j}).
\end{aligned}
\label{eq:augmented Q}
\end{equation}

We need the following assumption. \theoremstyle{definition}\begin{assumption}The
supplemental coefficient $\alpha_{i}$ satisfies $0\leq\alpha_{i+1}<\alpha_{i}<1,$
and $\lim_{i\rightarrow\infty}\alpha_{i}=0.$ $\mathcal{V}(x_{k})$
in \eqref{eq:augmented Q} is finite and positive definite in $x_{k}$.\end{assumption}Note
that the major difference between $Q^{(i)}(x_{k},u_{k})$ in \eqref{eq:augmented Q}
and the original $\check{Q}^{(i)}(x_{k},u_{k})$ in \eqref{eq:policy-evaluation-original}
is an extra term $\sum_{j=0}^{\infty}\alpha_{i}\mathcal{V}(x_{k+j})$.
In fact, \eqref{eq:augmented Q} can be rewritten as 
\begin{equation}
\begin{aligned}Q^{(i)}(x_{k},u_{k})= & \check{U}(x_{k},u_{k})+\alpha_{0}\mathcal{V}(x_{k})\\
 & +\sum_{j=1}^{\infty}[\check{U}(x_{k+j},h^{(i)}(x_{k+j}))+\alpha_{i}\mathcal{V}(x_{k+j})].
\end{aligned}
\label{eq:augmented Q 2}
\end{equation}
We can see that the stage cost $\check{U}(x_{k},u_{k})$ is supplemented
by $\mathcal{V}(x_{k})$. This augmented value $Q^{(i)}(x_{k},u_{k})$
is no longer the same as $\check{Q}^{(i)}(x_{k},u_{k})$ in the regular
formulation of policy iteration \eqref{eq:policy-evaluation-original}.
With this additional supplemental value, the learning agent receives
guidance to shape the sequence of stage costs $\check{U}$.

With such an augmented Q-value formulation in \eqref{eq:augmented Q 2},
the policy evaluation step \eqref{eq:policy-evaluation-original}
and policy improvement step \eqref{eq:policy-improvement-original}
becomes:

\subsubsection*{Policy Evaluation with Supplemental Value}

\begin{align}
Q^{(i)}(x_{k},u_{k})= & U^{(i)}(x_{k},u_{k})+Q^{(i)}(x_{k+1},h^{(i)}(x_{k+1}))\nonumber \\
 & ,i=0,1,2,\ldots\label{eq:policy-evaluation}
\end{align}
where $U^{(i)}(x_{k},u_{k})=\check{U}(x_{k},u_{k})+\alpha_{i}\mathcal{V}(x_{k}).$

\subsubsection*{Policy Improvement}

\begin{equation}
h^{(i+1)}(x_{k})=\underset{u_{k}}{\arg\min}Q^{(i)}(x_{k},u_{k}),i=0,1,2,\ldots\label{eq:policy-improvement}
\end{equation}

\begin{rem}
\label{rem:prior knowledge}The term $\mathcal{V}$ represents a supplemental
value obtained from a previous experiment where $\mathcal{V}(x_{k})=\underset{u_{k}}{\min}Q_{f}(x_{k},u_{k})$.
In the above, $Q_{f}(x_{k},u_{k})$ is a converged value function
from applying a naive FPI without any supplemental value $\mathcal{V}$
(Algorithm 1) or just PI. The supplemental value $\mathcal{V}(x_{k})$
so formulated from a previous experiment of a similar human subject
can capture essential information represented in the value $\mathcal{V}$
for state $x_{k}$. When this information is used in a new experiment,
the $Q^{(i)}(x_{k},u_{k})$ value has previously obtained information
embedded into the current learning. Note, however, that both experiments
must use the same stage cost and cost-to-go function constructs.
\end{rem}
Similar to regular policy evaluation \eqref{eq:policy-evaluation-original}
and \eqref{eq:policy-improvement-original}, we need the following
assumption for the initial control law $h^{(0)}(x)$ in \eqref{eq:policy-evaluation}
and \eqref{eq:policy-improvement}:\theoremstyle{definition}\begin{assumption}The
initial control law $h^{(0)}(x)$ is admissible.\end{assumption}
\begin{defn}
(Admissible Control \cite{Liu2014f}) : A control policy $h(x)$ is
admissible with respect to the value function $\check{Q}(x,u)$ \eqref{eq:augmented Q}
if $h(x)$ is continuous on $\mathcal{X}$, $h(0)=0$ and it stabilizes
system \eqref{eq:system}, and the corresponding value function $\check{Q}(x,u)$
\eqref{eq:augmented Q} is finite for $\forall x\in\mathcal{X}$.
\end{defn}
An initially feasible set of impedance control parameters are available
from the prosthesis manufacturers and/or trained technicians/experimenters
who can customize the prosthesis for individual patients. After all,
manual tuning of the impedance parameters is the current practice
in clinics. In Theorem 3, given an initially admissible control law
$h^{(0)}(x)$, we will show that the iterative control law $h^{(i)}(x)$
is also admissible for $i=1,2,\ldots$ . 

Solving \eqref{eq:policy-evaluation} and \eqref{eq:policy-improvement}
to obtain closed-form optimal solutions $Q^{*}(x_{k},u_{k})$ and
$h^{*}(x_{k})$ are difficult or nearly impossible. A value function
approximation (VFA) scheme replaces the exact value function in \eqref{eq:policy-evaluation}
with a function approximator such as neural networks. Such approximation
based approaches to solving the Bellman equation, or RL approaches,
usually utilize an actor-critic structure where the critic evaluates
the performance of a control policy and the actor improves the control
policy based on the critic evaluation. Both the actor and the critic
work together iteratively and learning takes place forward-in-time
to approximately solve the Bellman equation.

Our next strategy to improve policy evaluation efficiency is to innovatively
utilize experience replay. 

\subsection{Flexible Sampling with Experience Replay\label{subsec:Prioritized ER}}

In policy evaluation \eqref{eq:policy-evaluation}, the value function
$Q^{(i)}$ is to be evaluated with multiple samples of $s_{k}=(x_{k},u_{k},x_{k+1})$.
How many samples to use and how to select the samples directly impact
policy evaluation. We propose the following options to flexibly select
the number of samples and/or prioritize the samples in order to improve
policy evaluation.

Let $DS=\{s_{k}\}_{N}$ of size $N$ be a memory buffer. When there
is no abundance of data, it would be natural to perform a policy evaluation
of \eqref{eq:policy-evaluation} using not only newly available sample
but also all those samples already in the memory buffer $DS$ which
may include off-policy samples, on-policy samples, or both. 

Next, samples in $DS$ can be assigned with different priorities so
that the important samples are more likely to be reused. In this work,
the importance of sample $s_{k}$ is measured by the temporal difference
(TD) error from a transition \cite{Schaul2015}, which indicates how
surprising or unexpected the transition is: specifically, how far
the value is from its next-step bootstrap estimate.

Let $\delta_{k}^{(i)}$ be the TD error of sample $s_{k}$ in $DS$
under policy $h^{(i)}$. The rank $\zeta_{k}^{(i)}$ (ordinals from
1, which corresponds to the largest TD error) of sample $s_{k}$ is
obtained from sorting the memory buffer $DS$ according to $|\delta_{k}^{(i)}|$
in a descending order. Then each sample $s_{k}$ is assigned a weight
$\bar{\lambda}_{k}^{(i)}$ as 
\begin{equation}
\bar{\lambda}_{k}^{(i)}=\frac{1}{\zeta_{k}^{(i)}},\textrm{for }\forall k,\label{eq:unnormalized sample weight}
\end{equation}
and $\bar{\lambda}_{k}^{(i)}$ can be normalized to a value between
0 and 1 as 
\begin{equation}
\lambda_{k}^{(i)}=\frac{\bar{\lambda}_{k}^{(i)}}{\sum\bar{\lambda}_{k}^{(i)}},\textrm{for }\forall k.\label{eq:normalized sample weight}
\end{equation}
The $\lambda_{k}^{(i)}$ can then provide a flexibility for weighing
the samples when solving the Bellman equation. 

\subsection{Approximate Policy Evaluation with Flexibility \label{subsec:Approximate-Policy-Evaluation}}

To implement the policy evaluation step \eqref{eq:policy-evaluation},
a function approximator $\hat{Q}^{(i)}(x_{k},u_{k})$ is used for
$Q^{(i)}(x_{k},u_{k})$. We use a universal function approximator
such that: 

\begin{equation}
\hat{Q}^{(i)}(x_{k},u_{k})=W^{(i)T}\phi(x_{k},u_{k})=\sum_{j=1}^{L}w_{j}^{(i)}\varphi_{j}(x_{k},u_{k})\label{eq:Q_hat_W}
\end{equation}
where $W^{(i)}\in\mathbb{R}^{L}$ is a weight vector and $\phi(x_{k},u_{k}):\mathbb{R}^{N_{x}}\times\mathbb{R}^{N_{u}}\rightarrow\mathbb{R}^{L}$
is a vector of the basis functions $\varphi_{j}(x_{k},u_{k}),k=1\ldots L$.
The basis functions $\varphi_{j}(x_{k},u_{k})$ can be neural networks,
polynomial functions, radial basis functions, etc. In our implementations
(Section V), we employ polynomial basis, the associated universal
approximation property can be shown by the Stone-Weierstrass theorem.

The policy evaluation step \eqref{eq:policy-evaluation} then becomes
\begin{equation}
\begin{aligned} & \hat{Q}^{(i)}(x_{k},u_{k})\\
 & =U^{(i)}(x_{k},u_{k})+\hat{Q}^{(i)}(x_{k+1},h^{(i)}(x_{k+1})).
\end{aligned}
\label{eq:HJB-approximate}
\end{equation}
Substituting \eqref{eq:Q_hat_W} into \eqref{eq:HJB-approximate},
we have 
\begin{equation}
\begin{aligned}W^{(i)T}[\text{\ensuremath{\phi}}(x_{k},u_{k}) & -\text{\ensuremath{\phi}}(x_{k+1},h^{(i)}(x_{k+1}))]\\
 & =U^{(i)}(x_{k},u_{k}).
\end{aligned}
\label{eq:HJB-W}
\end{equation}
Equation \eqref{eq:HJB-approximate} can be seen as an approximated
policy evaluation step in terms of a weight vector that is to be determined
from solving $L$ linear equations. At iteration $i$, two column
vectors $X^{(i)}\in\mathbb{R}^{N\text{\texttimes}L}$ and $Y^{(i)}\in\mathbb{R}^{N}$,
are formed by the term $\text{\ensuremath{\phi}}(x_{k},u_{k})-\text{\ensuremath{\phi}}(x_{k+1},h^{(i)}(x_{k+1}))$
and $U^{(i)}(x_{k},u_{k})$, respectively, in each row. In other words,
\eqref{eq:HJB-W} can be rewritten as 
\begin{equation}
W^{(i)T}X^{(i)}=Y^{(i)}.\label{eq:sim_LS}
\end{equation}
The TD error $\delta_{k}^{(i)}$ can be computed as 
\begin{equation}
\begin{aligned}\delta_{k}^{(i)}= & U^{(i)}(x_{k},u_{k})+\hat{Q}^{(i-1)}(x_{k+1},h^{(i)}(x_{k+1}))\\
 & -\hat{Q}^{(i-1)}(x_{k},u_{k}),\textrm{for }i=0,1,2,\ldots
\end{aligned}
\label{eq:TD error}
\end{equation}
Then the weight $\lambda_{k}^{(i)}$ of a sample can be obtained from
\eqref{eq:normalized sample weight}. For $i=0$, assign weights $\lambda_{k}^{(0)}=1$.
When the policy evaluation with function approximation \eqref{eq:HJB-approximate}
is carried out with sample $s_{k}=(x_{k},u_{k},x_{k+1}),$ it can
be weighted by $\lambda_{k}^{(i)}$. Hence, the weight vector $W^{(i)}$
can be computed from \eqref{eq:sim_LS} as a weighted least squares
solution using $N$ weighted samples
\begin{equation}
W^{(i)}=(X^{(i)^{T}}\Lambda^{(i)}X^{(i)}){}^{\dagger}(X^{(i)^{T}}\Lambda^{(i)}Y^{(i)})^{T},\label{eq:W}
\end{equation}
where $\Lambda^{(i)}\text{\ensuremath{\in}}\mathbb{R}^{N}$ is a vector
of $\lambda_{k}^{(i)}$ and $^{\dagger}$ is the Moore-Penrose pseudoinverse.
Once $W^{(i)}$ is obtained, the approximated value function $\hat{Q}^{(i)}(x_{k},u_{k})$
can be obtained using \eqref{eq:Q_hat_W}. 

Note that, in \eqref{eq:sim_LS} to \eqref{eq:W} we use $N$ samples
to estimate the weight vector $W^{(i)}\in\mathbb{R}^{L}.$ For $\hat{Q}^{(i)}$
to be convergent, we need the following PE-like condition.\theoremstyle{definition}\begin{condition}The
vector $X^{(i)}$ formed by $N$ samples in $DS$ contains as many
linear independent elements as the unknown parameters in the weight
vector $W^{(i)}$, i.e., $rank(X^{(i)})=L$.\end{condition}
\begin{rem}
The number of samples $N$ in the memory buffer can be fixed or adaptive
with $N>L$ satisfied necessarily. Unlike PE, Condition 1 can be checked
in real time easily. Adding a small Guassian noise (for example, 1\%
of initial impedance value) to the impedance values $u_{k}=h^{(i)}(x_{k})$
suffices for meeting Condition 1.
\end{rem}
\begin{algorithm}[tbh]
\caption{Flexible Policy Iteration (FPI)\label{alg:1}}

\begin{raggedright}
Initialization by
\par\end{raggedright}
\begin{raggedright}
Random initial state $x_{0}\in\mathfrak{\mathcal{X}}$, initial batch
size $N_{b}$ (if in batch mode), memory buffer $DS=\emptyset$, initially
admissible control policy $h^{(0)}$. Let the approximated policy
$\hat{h}^{(0)}=h^{(0)}.$
\par\end{raggedright}
\begin{raggedright}
Data Preparation for Iteration $i$
\par\end{raggedright}
1a: (Batch Data Collection) Collect $N_{b}$ samples $\{(x_{k},u_{k},x_{k+1})\}_{N_{b}}$
from system \eqref{eq:system} following policy $\hat{h}^{(i)}$from
gait cycle $k$, $N\leftarrow N_{b}$ (Setting 2(A) in Table I).

1b: (Incremental Data Collection) Collect a sample $(x_{k},u_{k},x_{k+1})$
from system \eqref{eq:system} following policy $\hat{h}^{(i)}$,
and add it to $\ensuremath{DS}$, $N\leftarrow N+1$ (Setting 2(B)
in Table I).

2: (Set Batch Size) Either use a fixed or adaptive $N_{b}$ (Setting
1 in Table I) if under batch mode (Setting 2(A) in Table I).

3: (Set Other Parameters) Determine $\lambda_{k}^{(i)}$ (Setting
3 in Table I) and $\alpha_{i}$ (Setting 4 in Table I).

Policy Evaluation/Update for Iteration $i$

4: (Policy Evaluation) Evaluate policy $\hat{h}^{(i)}$ by solving
\eqref{eq:HJB-approximate} for $\hat{Q}^{(i)}$ using \eqref{eq:W},
for example, and by using all samples in $DS$.

5: (Policy Update) Update policy $\hat{h}^{(i+1)}$ by \eqref{eq:special policy 2}
and \eqref{eq:special policy 3}.
\end{algorithm}

\begin{table}[h]
\caption{Data Preparation and Parameter Settings in Algorithm 1\label{tab:Available-Options}}

\centering{}%
\begin{tabular}{lll}
\toprule 
\multicolumn{2}{l}{Setting} & Description\tabularnewline
\midrule
\multirow{2}{*}{1} & (A) $N_{b}$ is fixed & Fixed\tabularnewline
 & (B) $N_{b}\leftarrow N_{b}+5$ & Adaptive\tabularnewline
\midrule
\multirow{2}{*}{2} & (A) $N\leftarrow N_{b}$ & Batch mode\tabularnewline
 & (B) $N\leftarrow N+1$ & Incremental mode\tabularnewline
\midrule
\multirow{2}{*}{3} & (A) $\lambda_{k}^{(i)}=1$ & No prioritization\tabularnewline
 & (B) $\textrm{\ensuremath{\lambda_{k}^{(i)}} from \eqref{eq:normalized sample weight}}$ & With prioritization\tabularnewline
\cmidrule{2-3} \cmidrule{3-3} 
\multirow{2}{*}{4} & (A) $\alpha_{i}=0$ & No supplemental value\tabularnewline
 & (B) $\alpha_{i}=0.9^{i}$ & With supplemental value\tabularnewline
\bottomrule
\end{tabular}
\end{table}

\subsection{Policy Improvement in FPI\label{subsec:Implementation-of-Policy}}

After the approximated value function $\hat{Q}^{(i)}(x_{k},u_{k})$
is obtained, the next policy $h^{(i+1)}(x_{k})$ from \eqref{eq:policy-improvement}
is, 
\begin{equation}
h^{(i+1)}(x_{k})=\underset{u_{k}}{\arg\min}\hat{Q}^{(i)}(x_{k},u_{k}).\label{eq:policy-update}
\end{equation}

We employ another linear-in-parameter universal function approximator
$\hat{h}^{(i+1)}(x_{k})$ for $h^{(i+1)}(x_{k})$, 
\begin{equation}
\hat{h}^{(i+1)}(x_{k})=(\mathcal{K}^{(i+1)})^{T}\sigma(x_{k}),\label{eq:special policy 2}
\end{equation}
where $\mathcal{K}^{(i+1)}$ is a weight vector and $\sigma(x_{k})$
is a basis function vector. The weight vector $\mathcal{K}^{(i+1)}$
is updated iteratively using the gradient of the approximate value
function $\hat{Q}^{(i)}(x_{k},u_{k})$,

\begin{equation}
\mathcal{K}_{j+1}^{(i+1)}=\mathcal{K}_{j}^{(i+1)}-l\dfrac{\partial\hat{Q}^{(i)}(x_{k},(\mathcal{K}_{j}^{(i+1)})^{T}\sigma(x_{k}))}{\partial\mathcal{K}_{j}^{(i+1)}}\label{eq:special policy 3}
\end{equation}
where $l$ is the learning rate ($0<l<1$), $j$ is the iteration
step within a policy iteration step. 

\subsection{Implementation of FPI}

Algorithm \ref{alg:1} and Table I together describe our proposed
FPI algorithm. The terminating condition in Algorithm \ref{alg:1}
can be, for example, policy iteration index $i=i_{max}$ where $i_{max}$
is some positive number, or $|\hat{Q}^{(i)}(x_{k},u_{k})-\hat{Q}^{(i-1)}(x_{k},u_{k})|<\varepsilon$
where $\varepsilon$ is a small positive number. Note that there are
four settings in Algorithm 1 (Table I). FPI can run in batch mode
or incremental mode (Setting 2). In batch mode, only samples (of length
$N_{b}$) generated under the same policy are used in policy evaluation,
thus no sample reuse is allowed in this mode. In incremental mode,
previous samples of length $N$ in the $DS$ that are generated under
different policies, plus a newly acquired sample, can be reused to
evaluate a new policy. In batch mode, an extra parameter batch size
$N_{b}$ need to be set (Setting 1 in Table I), while such parameter
is not required under incremental mode. In addition, Setting 3 describes
how the priorities $\lambda_{k}^{(i)}$ of the samples are assigned
and Setting 4 describes how the supplemental value is used at iteration
$i$ through the parameter of $\alpha_{i}$ . 

Note that in batch mode, FPI can choose the number of samples for
policy evaluation adaptively. FPI starts with a small $N_{b}$. A
newly generated policy is tested with one or more gait cycles to determine
if the policy can lower the stage cost. If not, a larger set of samples
(e.g. $N_{b}\leftarrow N_{b}+5$) is used.

\begin{figure}[t]
\begin{centering}
\subfloat[Phase 1]{\includegraphics[width=0.45\columnwidth]{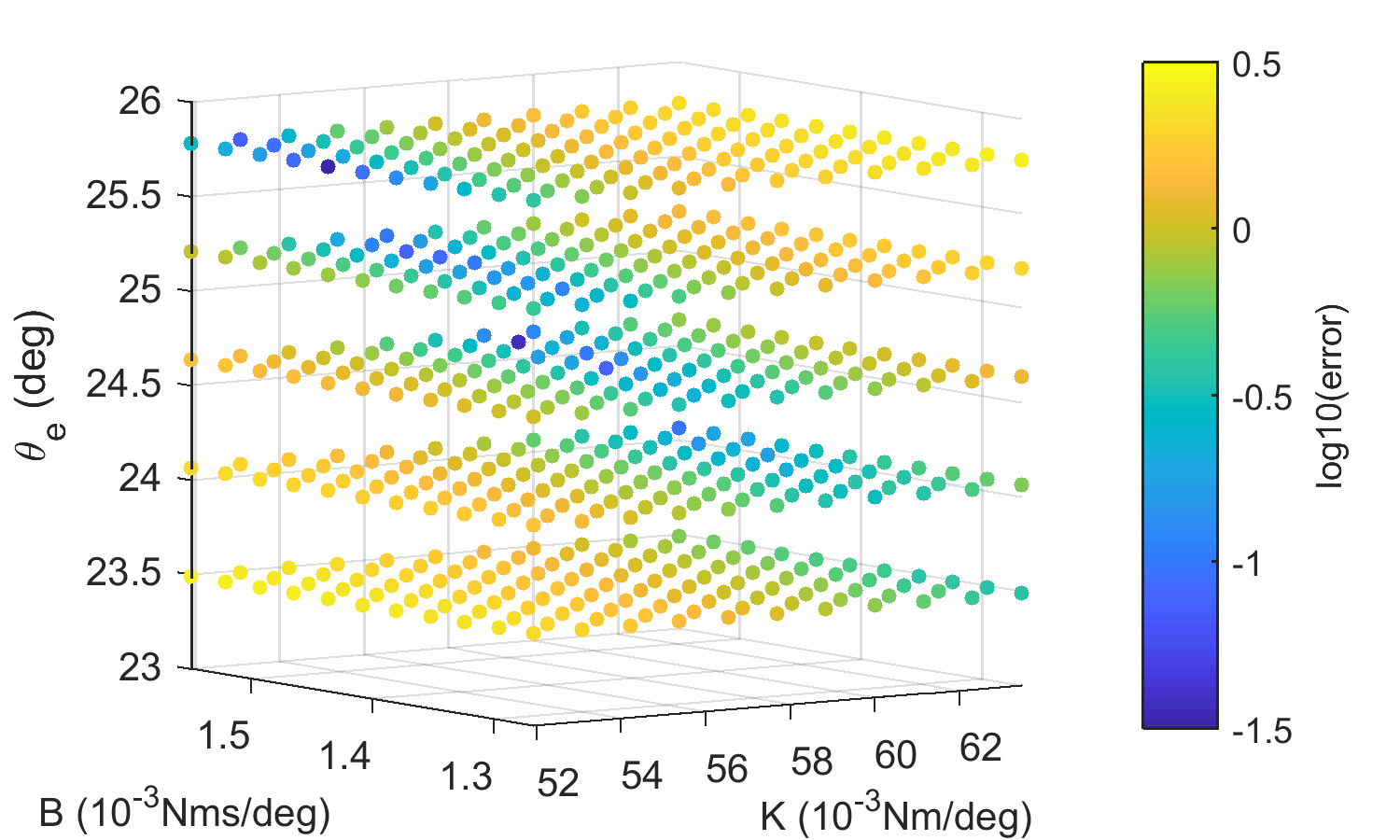}}\subfloat[Phase 2]{\includegraphics[width=0.45\columnwidth]{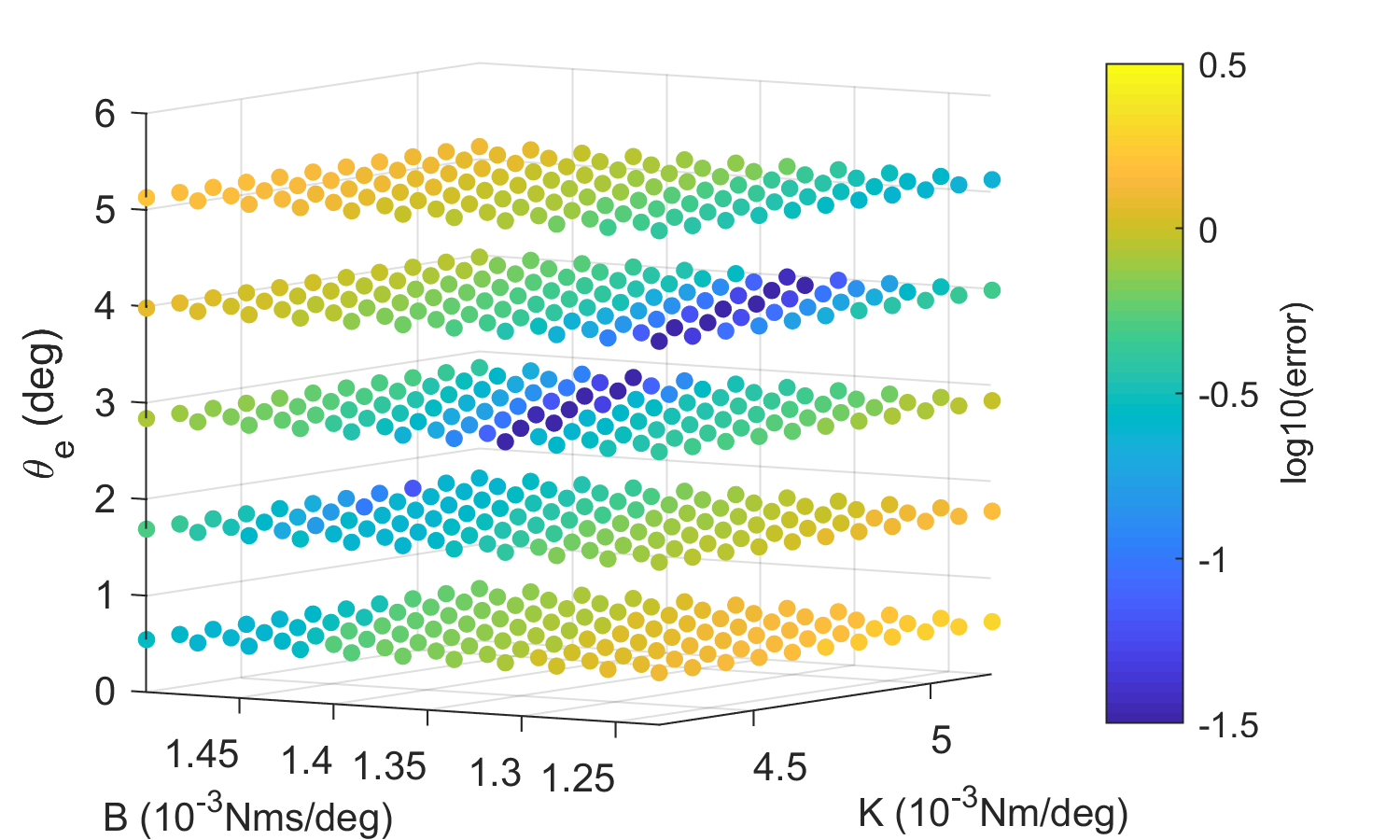}}\vspace{0in}
\subfloat[Phase 3]{\includegraphics[width=0.45\columnwidth]{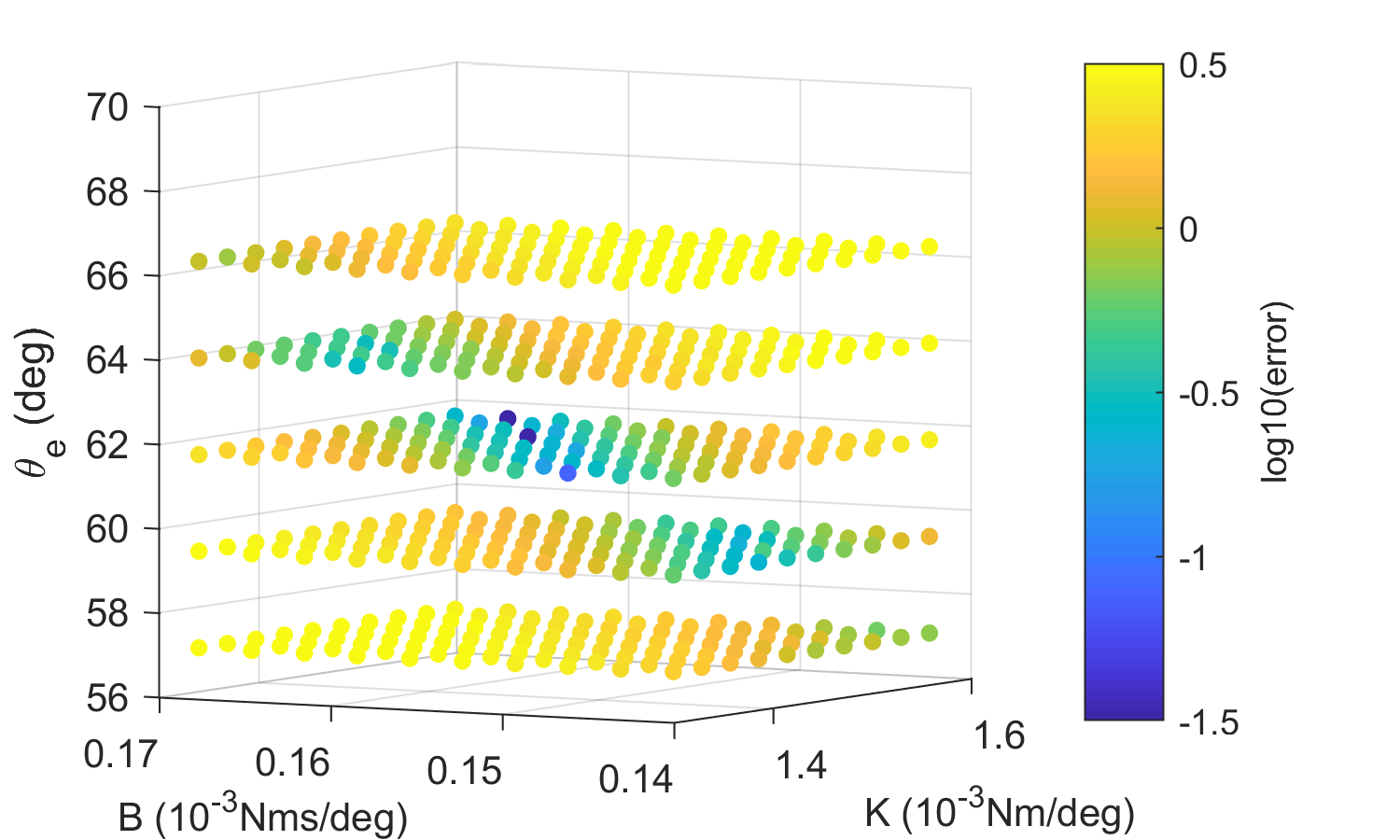}}\subfloat[Phase 4]{\includegraphics[width=0.45\columnwidth]{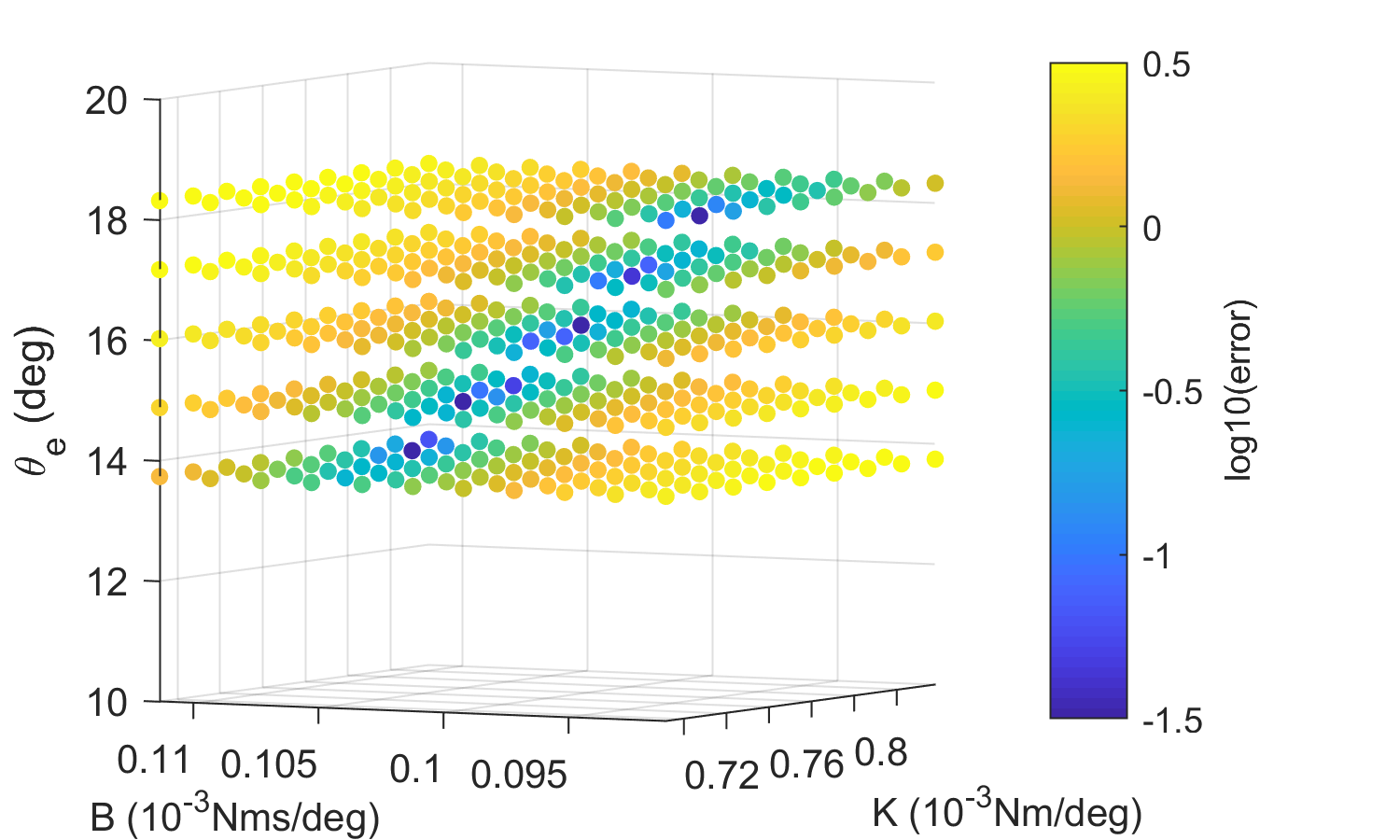}}
\par\end{centering}
\caption{The stage cost in peak error and duration error as in equation \eqref{eq:utility-function}.
\label{fig:Combined-Errors}}
\end{figure}

\begin{figure}[t]
\begin{centering}
\subfloat[Phase 1]{\includegraphics[width=0.45\columnwidth]{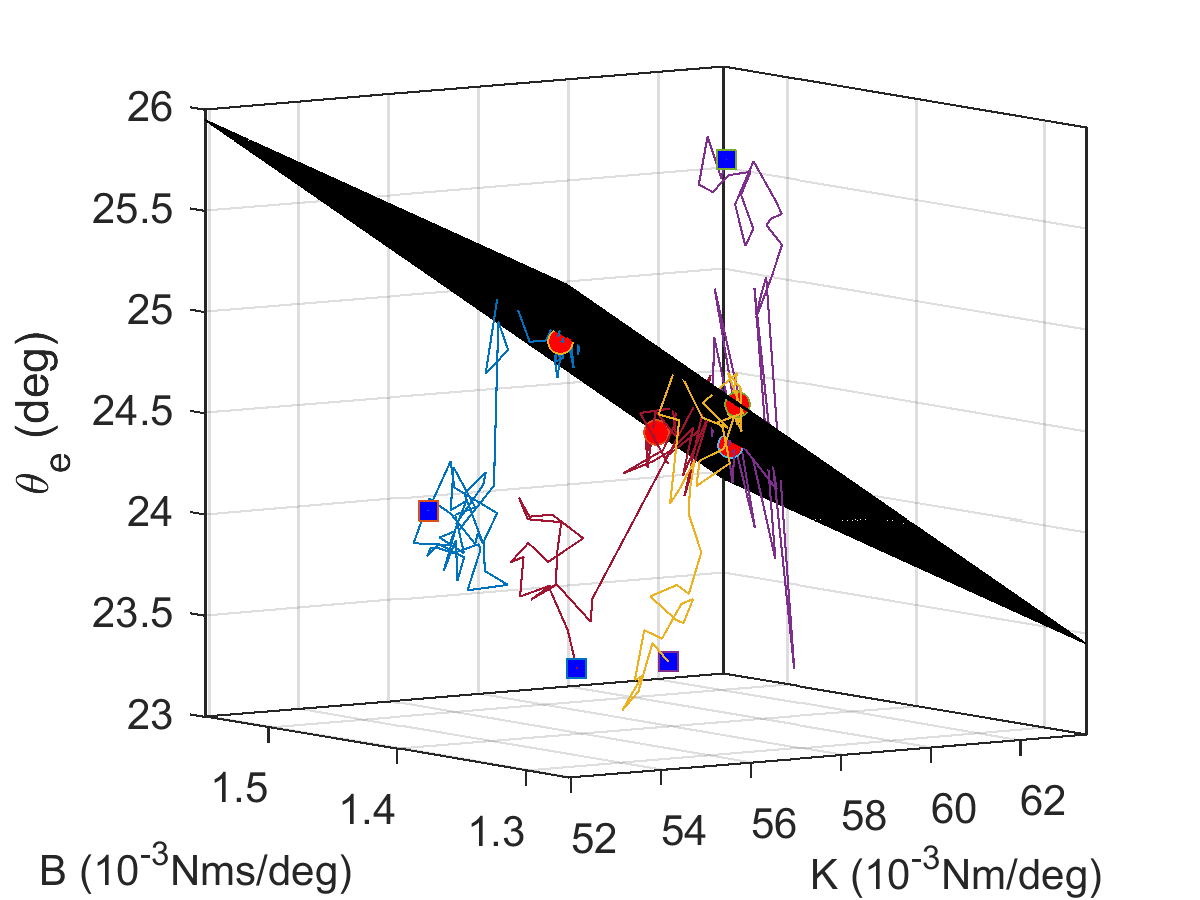}}\subfloat[Phase 2]{\includegraphics[width=0.45\columnwidth]{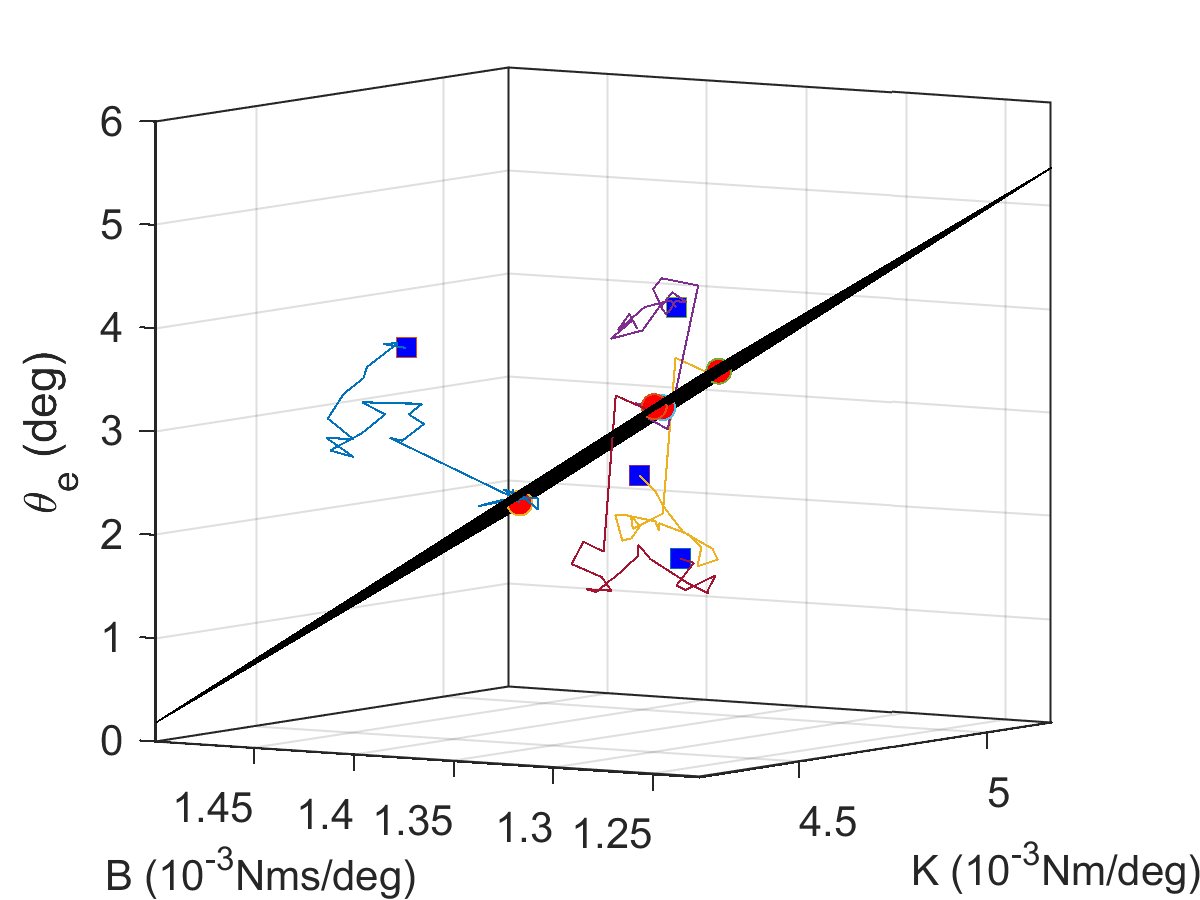}}\vspace{0in}
\subfloat[Phase 3]{\includegraphics[width=0.45\columnwidth]{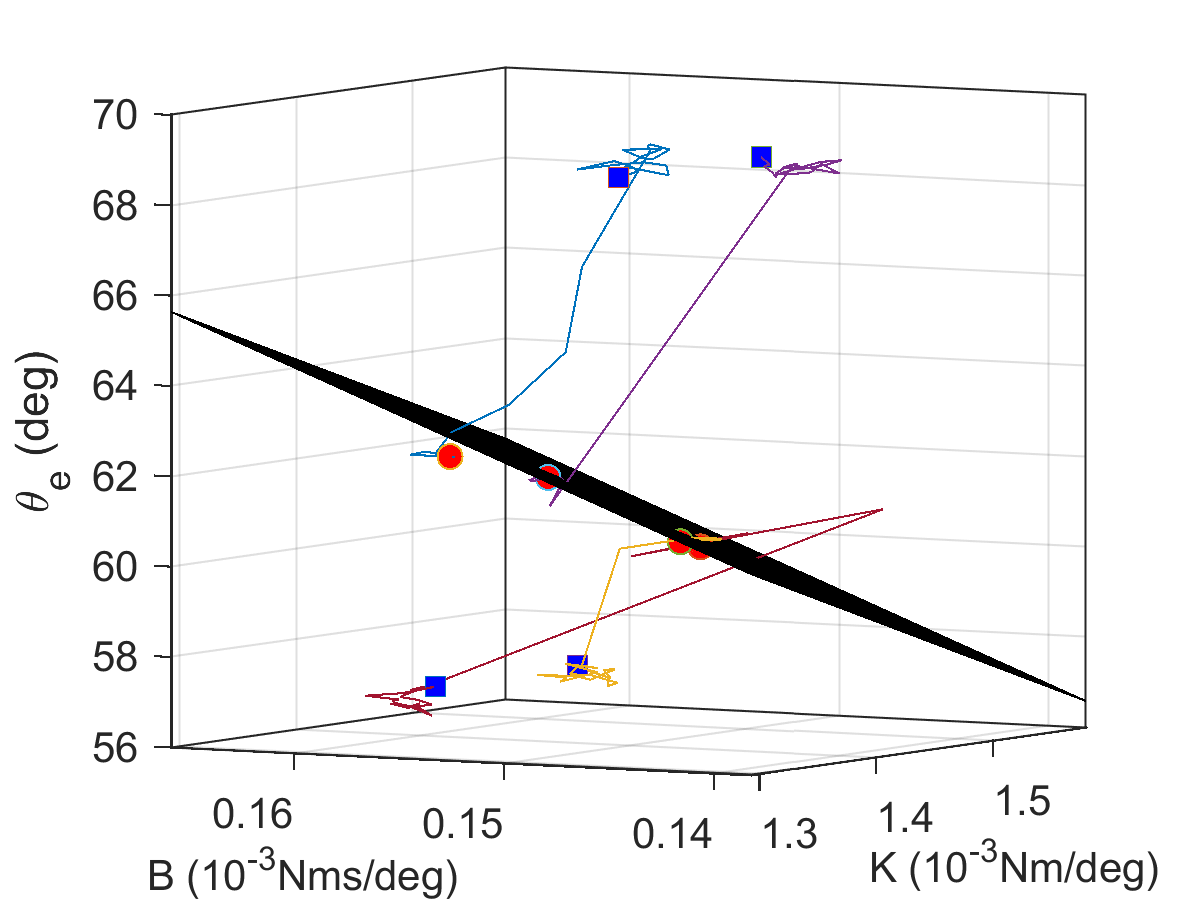}}\subfloat[Phase 4]{\includegraphics[width=0.45\columnwidth]{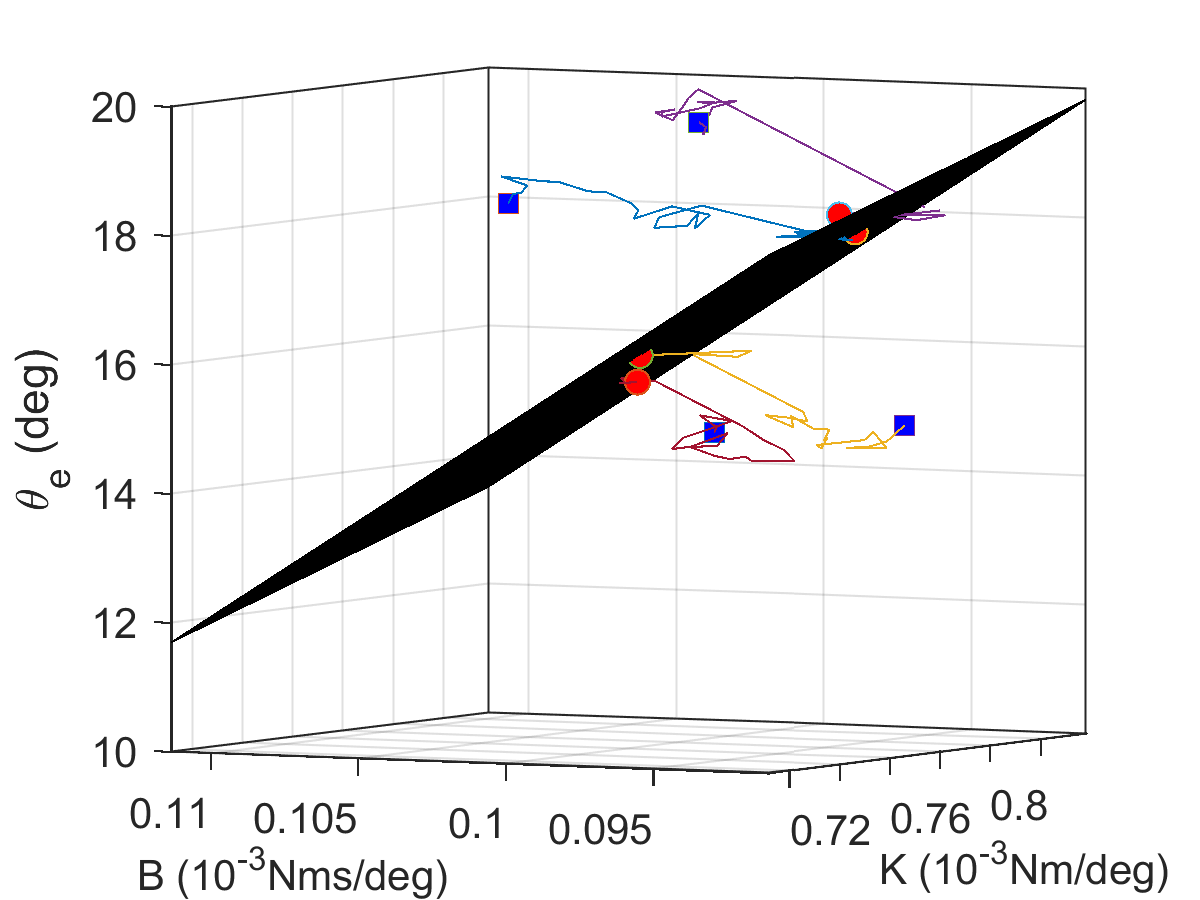}}
\par\end{centering}
\caption{Illustration of the converging process of the IC parameters during
FPI tuning: from randomly initialized IC parameters (four trials for
illustration here, shown in blue squares) to the final parameters
(shown in red dots), which are fitted with a regression response surface.\label{fig:Response-Manifold}}
\end{figure}

This adaptive approach is based on our observations as follows. Given
a continuous state and control problem such as the control of a robotic
knee, we constructed a quadratic stage cost $(x_{k},u_{k})$ in \eqref{eq:utility-function}
which is common in control system design. As a decreasing stage cost
can be viewed as necessary toward an improved value during each iteration,
it thus becomes a natural choice for such a selection criterion. For
example, Fig. \ref{fig:Combined-Errors} depicts stage cost for the
uniformly sampled IC parameter space in our human-robot application,
where the color of each sample point represents a stage cost. Fig.
\ref{fig:Response-Manifold} was generated under the setting of (A)(A)(A)(A)
in Table \ref{tab:Available-Options} and $N_{b}=20$. Fig. \ref{fig:Response-Manifold}
shows the trajectories of the IC parameters tuned by FPI starting
from some random initial IC parameters. Apparently, the points with
minimum stage cost in Fig. \ref{fig:Combined-Errors} coincides with
the converging planes found by FPI in Fig. \ref{fig:Response-Manifold}.

\section{Qualitative Properties of FPI\label{sec:Properties-of-FPI}}

Policy iteration based RL has been shown possessing several important
properties, such as convergence of policy iteration, approximately
reaching Bellman optimality, and stabilizing control \cite{Mu2017},
\cite{Liu2014f}-\cite{Guo2017}.  We now address the question of
whether these properties still hold under our proposed FPI framework
especially when our formulation includes a supplemental value. 
\begin{lem}
Let $i=0,1,...$ be the iteration number and let $Q^{(i)}(x_{k},u_{k})$
and $h^{(i)}(x_{k})$ be updated by \eqref{eq:policy-evaluation}-\eqref{eq:policy-improvement}.
Under Assumption 1, the stage cost $U^{(i)}(x_{k},u_{k})$ and iterative
value function $Q^{(i)}(x_{k},u_{k})$ in \eqref{eq:policy-evaluation}
are positive definite for $x_{k}$ and $u_{k}$.
\end{lem}
\begin{IEEEproof}
For $i=0$, according to Assumption 1, we have $h^{(0)}(x_{k})=0$
as $x_{k}=0$. As $\check{U}(x_{k},u_{k})$ is positive definite for
$x_{k}$ and $u_{k}$, we have that $\sum_{j=0}^{\infty}\check{U}(x_{k+j},h^{(0)}(x_{k+j}))=0$
as $x_{k}=0,$ and $\sum_{j=0}^{\infty}\check{U}(x_{k+j},h^{(0)}(x_{k+j}))>0$
for any $x_{k}\neq0$. Hence $\sum_{j=0}^{\infty}\check{U}(x_{k+j},h^{(0)}(x_{k+j}))$
is a positive definite function for $x_{k}$. Since $\mathcal{V}(x_{k})$
is also positive definite for $x_{k}$, it is easy to get $U^{(i)}(x_{k},u_{k})$
is positive definite for $x_{k}$ and $u_{k}$. According to \eqref{eq:augmented Q},
if $x_{k}=u_{k}=0$, $Q^{(0)}(x_{k},u_{k})=0$; if $|x_{k}|+|u_{k}|\neq0$,
$Q^{(0)}(x_{k},u_{k})>0$, which proves that $Q^{(0)}(x_{k},u_{k})$
is positive definite for $x_{k}$ and $u_{k}.$ Based on this idea,
we can prove that the iterative function $Q^{(i)}(x_{k},u_{k}),i=0,1,...,$
is positive definite for $x_{k}$ and $u_{k}.$
\end{IEEEproof}

\begin{thm}
Let Assumptions 1-3 hold. Let $Q^{(i)}(x_{k},u_{k})$ and $h^{(i)}$
be updated by \eqref{eq:policy-evaluation}-\eqref{eq:policy-improvement}.
Then, for $i=0,1,2,...,$ $h^{(i)}$ stabilizes the system \eqref{eq:system}.
\end{thm}
\begin{IEEEproof}
Consider the case when $x_{k}\neq0$, we have $U^{(i)}(x_{k},h^{(i)}(x_{k}))>0$
and $\alpha_{i}(x_{k+1})\geq0.$ From \eqref{eq:policy-evaluation},
and $i=0,1,\ldots,$
\begin{equation}
\begin{aligned} & Q^{(i)}(x_{k},h^{(i)}(x_{k}))-Q^{(i)}(x_{k+1},h^{(i)}(x_{k+1}))\\
 & =U^{(i)}(x_{k},h^{(i)}(x_{k}))>0.
\end{aligned}
\end{equation}
Next, consider the case when $x_{k}=0$, according to Assumption 1
we can get $h^{(i)}(x_{k})=0$ and thus $U^{(i)}(x_{k},h^{(i)}(x_{k}))=0$,
which imply $Q^{(i)}(x_{k},h^{(i)}(x_{k}))-Q^{(i)}(x_{k+1},h^{(i)}(x_{k+1}))=0.$
According to Lemma 1 and Assumption 1, the function $Q^{(i)}(x_{k},h^{(i)}(x_{k}))$
is positive definite for $x_{k}$. Then $Q^{(i)}(x_{k},h^{(i)}(x_{k}))$
is a Lyapunov function. Thus $h^{(i)}$ stabilizes the system \eqref{eq:system}.
\end{IEEEproof}
\begin{rem}
Theorem 1 shows that the Lyapunov stability can be guaranteed under
iterative policy $h^{(i)}(x_{k})$ under the augmented value formulation
of \eqref{eq:augmented Q}. Additionally, embedded safety constraints
on the knee joint angles and angular velocities \cite{Wen2019} ensure
the states and the controls are within the domain $D$ of the system
dynamics $F(x_{k},u_{k})$ in \eqref{eq:system}. Human subjects are
therefore guaranteed to be practically stable.
\end{rem}
\begin{thm}
Let Assumptions 1-3 hold. Let the value function $Q^{(i)}(x_{k},u_{k})$
and the control policy $h^{(i)}(x_{k})$ be obtained from \eqref{eq:policy-evaluation}
and \eqref{eq:policy-improvement}, respectively. Then $Q^{(i+1)}(x_{k},u_{k})\leq Q^{(i)}(x_{k},u_{k})$
holds for $i=0,1,2,\ldots$ and $\forall(x_{k},u_{k})\in\mathcal{D}.$
\end{thm}
\begin{IEEEproof}
For convenience, we will use the following short hand notations in
the derivations, e.g., $U^{(i)}(x_{k},h^{(i)})$ for $U^{(i)}(x_{k},h^{(i)}(x_{k}))$.
According to \eqref{eq:augmented Q}, we can define $V(x_{k})$ as

\begin{equation}
V^{(i)}(x_{k})=Q^{(i)}(x_{k},h^{(i)})=\sum_{j=k}^{\infty}U^{(i)}(x_{j},h^{(i)}).\label{eq:V value function}
\end{equation}
Based on \eqref{eq:policy-improvement}, we have
\begin{align}
Q^{(i)}(x_{k},h^{(i+1)}) & =\underset{u_{k}}{\min}Q^{(i)}(x_{k},u_{k})\nonumber \\
 & \leq Q^{(i)}(x_{k},h^{(i)}).\label{eq:proof 2}
\end{align}
Based on \eqref{eq:policy-evaluation} we have 
\begin{align*}
V^{(i)}(x_{k}) & =Q^{(i)}(x_{k},h^{(i)})\\
 & \geq Q^{(i)}(x_{k},h^{(i+1)})\\
 & =U^{(i)}(x_{k},h^{(i+1)})+V^{(i)}(x_{k+1})
\end{align*}
\begin{equation}
\begin{aligned} & \geq U^{(i)}(x_{k},h^{(i+1)})+V^{(i)}(x_{k+1}).\end{aligned}
\end{equation}
Hence
\begin{eqnarray*}
V^{(i)}(x_{k})-V^{(i)}(x_{k+1}) & \geq & U^{(i)}(x_{k},h^{(i+1)})
\end{eqnarray*}
\begin{align}
V^{(i)}(x_{k+1})- & V^{(i)}(x_{k+2})\nonumber \\
 & \geq U^{(i)}(x_{k+1},h^{(i+1)})\nonumber \\
 & \vdots\nonumber \\
V^{(i)}(x_{k+L})- & V^{(i)}(x_{k+L+1})\nonumber \\
 & \geq U^{(i)}(x_{k+L},h^{(i+1)}).\label{eq:sumsum}
\end{align}
Summing up the left and the right hand sides of \eqref{eq:sumsum}
respectively,
\begin{equation}
\begin{aligned}V^{(i)}(x_{k})- & V^{(i)}(x_{k+L+1})\\
 & \geq\sum_{j=k}^{k+L}U^{(i)}(x_{j},h^{(i+1)})\\
 & \geq0,
\end{aligned}
\label{eq:Qi+1 and Qi}
\end{equation}
where $L$ is a positive integer corresponding to gait cycles in this
paper. Hence, \eqref{eq:Qi+1 and Qi} yields 
\begin{equation}
V^{(i)}(x_{k})\geq V^{(i+1)}(x_{k}).\label{eq:Qi+1 and Qi - 2}
\end{equation}
According to \eqref{eq:policy-evaluation} and \eqref{eq:Qi+1 and Qi - 2},
we can obtain 
\begin{align}
Q^{(i+1)}(x_{k},u_{k}) & =U^{(i)}(x_{k},u_{k})+V^{(i+1)}(x_{k+1})\nonumber \\
 & \leq U^{(i)}(x_{k},u_{k})+V^{(i)}(x_{k+1})\nonumber \\
 & =Q^{(i)}(x_{k},u_{k}).
\end{align}
\end{IEEEproof}
\begin{thm}
Let Assumptions 1-3 hold. Let $Q^{(i)}(x_{k},u_{k})$ and $h^{(i)}$
be updated by \eqref{eq:policy-evaluation}-\eqref{eq:policy-improvement},
respectively. Then for $i=0,1,2,\ldots,$ $h^{(i)}$ is an admissible
control policy.
\end{thm}
\begin{IEEEproof}
From \eqref{eq:augmented Q} and Theorem 2 we have 
\begin{equation}
\begin{aligned}Q^{(0)}(x_{k},u_{k}) & \geq Q^{(1)}(x_{k},u_{k})\\
 & =U^{(1)}(x_{k},u_{k})+\sum_{j=1}^{\infty}U^{(1)}(x_{k+j},h^{(1)}(x_{k+j}))
\end{aligned}
\end{equation}
As $Q^{(0)}(x_{k},u_{k})$ is finite given $h^{(0)}$ is admissible
for $x_{k}$, $u_{k}$, we have $Q^{(1)}(x_{k},u_{k})$ is also finite
for $x_{k}$, $u_{k}$, and thus $\sum_{j=1}^{\infty}U^{(1)}(x_{k+j},h^{(1)}(x_{k+j}))<\infty$.
Given Assumption 1 and Theorem 1, we can conclude that $h^{(1)}$
is admissible. By mathematical induction, we can prove $h^{(i)}$
is admissible for $i=0,1,2,\ldots.$ 
\end{IEEEproof}
\begin{thm}
Let Assumptions 1-3 hold. Let the iterative value function $Q^{(i)}(x_{k},u_{k})$
and the control policy $h^{(i)}(x_{k})$ be obtained from \eqref{eq:policy-evaluation}
and \eqref{eq:policy-improvement}, respectively, and the optimal
value function $Q^{*}(x_{k},u_{k})$ and the optimal policy be defined
in \eqref{eq:Bellman optimality} and \eqref{eq:optimal_policy},
respectively. Then $Q^{(i)}(x_{k},u_{k})\rightarrow Q^{*}(x_{k},u_{k})$
and $h^{(i)}(x_{k})\rightarrow h^{*}(x_{k})$ as $i\rightarrow\infty$,
$\forall(x_{k},u_{k})\in\mathcal{D}$.
\end{thm}
\begin{IEEEproof}
By definition, $Q^{*}(x_{k},u_{k})\leq Q^{(i)}(x_{k},u_{k})$ holds
for any $i$, and from Theorem 2, $\left\{ Q^{(i)}(x_{k},u_{k})\right\} $
is a non-increasing sequence that is bounded by $Q^{*}(x_{k},u_{k})$.
Hence $\left\{ Q^{(i)}(x_{k},u_{k})\right\} $ must have a limit as
$i\rightarrow\infty.$ Denote this limit as $Q^{(\infty)}(x_{k},u_{k})\triangleq\lim_{i\rightarrow\infty}Q^{(i)}(x_{k},u_{k})$
and $h^{(\infty)}(x_{k})\triangleq\lim_{i\rightarrow\infty}h^{(i)}(x_{k})$.
Note that $\lim_{i\rightarrow\infty}U^{(\infty)}(x_{k},u_{k})=\check{U}(x_{k},u_{k}),$
take the limits in \eqref{eq:policy-evaluation} and \eqref{eq:policy-improvement}
as $i\rightarrow\infty,$
\begin{equation}
Q^{(\infty)}(x_{k},u_{k})=\check{U}(x_{k},u_{k})+Q^{(\infty)}(x_{k+1},h^{(\infty)}(x_{k})),\label{eq:Qinf}
\end{equation}

\begin{equation}
h^{(\infty)}(x_{k})=\underset{u_{k}}{\arg\min}Q^{(\infty)}(x_{k},u_{k}).\label{eq:hinf}
\end{equation}
The Bellman optimality equation for $V(x_{k})$ is 
\begin{equation}
V^{*}(x_{k})=\underset{h(.)}{\min}\left[\check{U}(x_{k},h(x_{k}))+V^{*}(x_{k+1})\right].\label{eq:V function Bellman optimality}
\end{equation}
When $i\rightarrow\infty$, $u_{k}=h^{(\infty)}(x_{k}),$ so from
\eqref{eq:Qinf} and \eqref{eq:hinf} we can get
\begin{eqnarray}
V^{(\infty)}(x_{k}) & = & Q^{(\infty)}(x_{k},h^{(\infty)}(x_{k}))\nonumber \\
 & = & \underset{u_{k}}{\min}\left[\check{U}(x_{k},u_{k})+Q^{(\infty)}(x_{k+1},h^{(\infty)}(x_{k}))\right]\nonumber \\
 & = & \underset{u_{k}}{\min}\left[\check{U}(x_{k},u_{k})+V^{(\infty)}(x_{k+1})\right].\label{eq:V inf}
\end{eqnarray}
Equation \eqref{eq:V inf} satisfies the Bellman optimality equation
\eqref{eq:V function Bellman optimality}, thus $V^{(\infty)}(x_{k})=V^{*}(x_{k})$.
From \eqref{eq:Qinf} we can obtain
\begin{align}
Q^{(\infty)}(x_{k},u_{k}) & =\check{U}(x_{k},u_{k})+V^{(\infty)}(x_{k+1})\nonumber \\
 & =\check{U}(x_{k},u_{k})+V^{*}(x_{k+1})\nonumber \\
 & =Q^{*}(x_{k},u_{k}).
\end{align}
Therefore $h^{(\infty)}(x_{k})=h^{*}(x_{k})$ can be obtained from
\eqref{eq:hinf}.
\end{IEEEproof}
Next, we consider the case of different types of errors that may affect
the Q-function, such as value function approximation errors, policy
approximation errors and errors from using $N$ samples to evaluate
the $i$th policy during policy iteration. We show an error bound
analysis of FPI while taking into account approximation errors.

We need the following assumption to proceed.\theoremstyle{definition}\begin{assumption}There
exists a finite positive constant $\gamma$ that makes the condition
$\underset{u_{k+1}}{\min}Q^{*}(x_{k+1},u_{k+1})\leq\gamma\check{U}(x_{k},u_{k})$
hold uniformly on $\mathfrak{\mathcal{X}}$.\end{assumption}For most
nonlinear systems, it is easy to find a sufficiently large number
$\gamma$ to satisfy this assumption as $Q^{*}(\cdotp)$ and $U(\cdotp)$
are finite.

Define a value function $\bar{Q}^{(i)}$ as 
\[
\bar{Q}^{(i)}(x_{k},u_{k})=\check{U}(x_{k},u_{k})+\hat{Q}^{(i-1)}(x_{k+1},h^{(i)}(x_{k+1}))
\]
 for $i=1,2,\ldots$ and $\bar{Q}^{(0)}=Q^{(0)}.$ Given the existence
of universal approximators, the total approximation error can be considered
finite during a single iteration, and therefore
\begin{equation}
\xi Q^{(i)}\leq\hat{Q}^{(i)}\leq\eta\bar{Q}^{(i)}\label{eq:error condition}
\end{equation}
holds uniformly for $i$ as well as $x_{k}$ and $u_{k}$, where $0<\xi\leq1$
and $\eta\geq1$ are constants, $\hat{Q}^{(i)}(x_{k},u_{k})$ is defined
by \eqref{eq:HJB-approximate} and $Q^{(i)}$ is defined by \eqref{eq:augmented Q}. 
\begin{thm}
Let Assumptions 1-4 hold. Let $\hat{Q}^{(i)}(x_{k},u_{k})$ be defined
by \eqref{eq:HJB-approximate} and $Q^{(i)}$ be defined by \eqref{eq:augmented Q}.
Given $1\leq\beta<\infty$ that makes $Q^{*}\leq Q^{(0)}\leq\beta Q^{*}$
hold uniformly for $x_{k},u_{k}.$ Let the approximate Q-function
$\hat{Q}^{(i)}$ satisfies the iterative error condition \eqref{eq:error condition}.
If the following condition is satisfied 
\begin{equation}
\eta<\frac{\gamma+1}{\gamma},\label{eq:theorem approx condition}
\end{equation}
then the iterative approximate Q-function $\hat{Q}^{(i)}$ is bounded
by 
\begin{equation}
\begin{aligned} & \xi Q^{*}\leq\hat{Q}^{(i)}\\
 & \leq\left[\eta\beta(\frac{\eta\gamma}{1+\gamma})^{i}+(1-(\frac{\eta\gamma}{1+\gamma})^{i})\frac{\eta}{1+\gamma-\eta\gamma}\right]Q^{*}.
\end{aligned}
\label{eq:theorem approx error long}
\end{equation}
Moreover, as $i\rightarrow\infty$, the approximate Q-function sequence
$\{\hat{Q}^{(i)}\}$ approaches $Q^{*}$ bounded by: 
\begin{equation}
\xi Q^{*}\leq\hat{Q}^{(\infty)}\leq\frac{\eta}{1+\gamma-\eta\gamma}Q^{*}.\label{eq:theorem approx error}
\end{equation}
\end{thm}
\begin{IEEEproof}
The left-hand side of \eqref{eq:theorem approx error long} can be
easily obtained according to \eqref{eq:error condition} and Theorem
3. 

The right-hand side of \eqref{eq:theorem approx error long} is proven
by mathematical induction as follows. 

First, for $i=0$, $\hat{Q}^{(0)}\leq\eta\bar{Q}^{(0)}=\eta Q^{(0)}\leq\eta\beta Q^{*}$
holds according to \eqref{eq:error condition} and the conditions
in Theorem 5. Thus \eqref{eq:theorem approx error long} holds for
$i=0$. 

Assuming that \eqref{eq:theorem approx error long} holds for $i\geq0$,
then for $i+1$ we have 
\begin{equation}
\begin{aligned} & \bar{Q}^{(i+1)}(x_{k},u_{k})\\
 & =U^{(i)}(x_{k},u_{k})+\hat{Q}^{(i)}(x_{k+1},h^{(i+1)}(x_{k+1}))\\
 & =U^{(i)}(x_{k},u_{k})+\underset{u_{k+1}}{\min}\hat{Q}^{(i)}(x_{k+1},u_{k+1})\\
 & \leq U^{(i)}(x_{k},u_{k})+\underset{u_{k+1}}{\min}P_{i}Q^{*}(x_{k+1},u_{k+1}),
\end{aligned}
\label{eq:approx error proof 1}
\end{equation}
where 
\begin{equation}
P_{i}=\eta\beta(\frac{\eta\gamma}{1+\gamma})^{i}+(1-(\frac{\eta\gamma}{1+\gamma})^{i})\frac{\eta}{1+\gamma-\eta\gamma}.
\end{equation}
According to Assumption 4, \eqref{eq:approx error proof 1} yields
\[
\begin{aligned} & \bar{Q}^{(i+1)}(x_{k},u_{k})\\
 & \leq(1+\gamma\frac{P_{i}-1}{\gamma+1})U^{(i)}(x_{k},u_{k})\\
 & \qquad+(P_{i}-\frac{P_{i}-1}{\gamma+1})\underset{u_{k+1}}{\min}Q^{*}(x_{k+1},u_{k+1})\\
 & =\frac{1}{\eta}\left[\eta\beta(\frac{\eta\gamma}{1+\gamma})^{i+1}+(1-(\frac{\eta\gamma}{1+\gamma})^{i+1})\frac{\eta}{1+\gamma-\eta\gamma}\right]\\
 & \qquad\left[U^{(i)}(x_{k},u_{k})+\underset{u_{k+1}}{\min}Q^{*}(x_{k+1},u_{k+1})\right]
\end{aligned}
\]
\begin{equation}
\begin{aligned} & =\frac{1}{\eta}\left[\eta\beta(\frac{\eta\gamma}{1+\gamma})^{i+1}+(1-(\frac{\eta\gamma}{1+\gamma})^{i+1})\frac{\eta}{1+\gamma-\eta\gamma}\right]\\
 & \qquad\times Q^{*}(x_{k},u_{k}).
\end{aligned}
\end{equation}
On the other hand, according to \eqref{eq:error condition}, there
is $\hat{Q}^{(i+1)}\leq\eta\bar{Q}^{(i+1)}$. Thus \eqref{eq:theorem approx error long}
holds for $i+1.$ By mathematical induction, the proof for \eqref{eq:theorem approx error long}
is completed. 

Considering \eqref{eq:error condition} and \eqref{eq:theorem approx error long},
we can easily obtain 
\begin{equation}
\hat{Q}^{(\infty)}\leq\frac{\eta}{1+\gamma-\eta\gamma}Q^{*}
\end{equation}
as $i\rightarrow\infty.$ Thus \eqref{eq:theorem approx error} holds. 
\end{IEEEproof}
\begin{rem}
Condition \eqref{eq:theorem approx condition} ensures that the upper
bound in \eqref{eq:theorem approx error} is finite and positive.
When $\xi=1$ and $\eta=1$, there is $Q^{*}\leq\hat{Q}^{(\infty)}\leq Q^{*}$
according to Theorem 5. Hence, $\hat{Q}^{(\infty)}=Q^{*}$. This means
when $\xi=1$ and $\eta=1$, the sequence of $\hat{Q}^{(i)}$ converges
to $Q^{*}$ as $i\rightarrow\infty$.
\end{rem}

\section{Robotic Knee Impedance Control By FPI\label{sec:Results}}

We are now in a position to apply FPI to solving the robotic knee
impedance control parameter tuning problem that originally motivated
our development of the FPI. Refer to Fig. 1. At the start of a gait
cycle, an initially feasible set of impedance parameters as those
in \eqref{eq:imp} are applied to FS-IC (top of Fig. 1(a), the IC
loop) so that OpenSim can provide a simulated knee angle dynamic trajectory
of a complete gait cycle including four gait phases (Fig. 1(b)). States
as in \eqref{eq:ADP-state} are then obtained for each of the 4 phases.
The actor and critic networks are initialized with random weights,
which are updated iteratively by \eqref{eq:policy-evaluation} and
\eqref{eq:policy-improvement} according to Algorithm 1 while data
preparation and FPI parameter settings are specified by the designer
according to Table I. Control policy from each iteration is used to
update the impedance parameter setting as in \eqref{eq:imp-update}
and \eqref{eq:u_k}, which in turn, result in control torques (2)
that interact with the FS-IC. The next gait cycle repeats the same
stimulation process. Note that the four RL controllers are of the
same structure (i.e., there are four independent FPI blocks in Fig.
1 but the initial state of phase $i+1$ is the end state of phase
$i,i=1,2,3$). 

We used OpenSim (https://simtk.org/) to simulate the dynamics of the
human-prosthesis system. OpenSim is a widely accepted simulator of
human movements. To simulate walking patterns of a unilateral above-knee
amputee, the right knee was treated as a prosthetic knee and controlled
by FS-IC, while the other joints in the model (left hip, right hip
and left knee) were set to follow prescribed motions.

The OpenSim walking model is deterministic, various noise patterns
(including real noise from human measurements) were added into the
simulations to realistically reflect a human-robot system. In Subsection
\ref{subsec:Comparison-Study}, noise was either generated by a random
number generator (the sensor noise and actuator noise cases in Table
\ref{tab:Comparison}), or by gait-to-gait variances captured from
two amputee subjects walking with prosthesis (case TF1 and TF2 in
Table \ref{tab:Comparison}). For the latter case, data were collected
from another study \cite{Brandt2017} where the experiments were approved
by the Institutional Review Board at the University of North Carolina
at Chapel Hill. To apply real gait-to-gait variance in simulation,
a total of 120 gait cycles of the intact knee joint movement trajectories
were used to compute deviations from the average joint motions and
then applied to the prescribed joint motions in the OpenSim model
accordingly. 

\subsection{Algorithm and Experiment Settings \label{subsec:Algorithm-and-Experiment} }

We summarize the parameters of the FPI in OpenSim simulations as follows.
Algorithm \ref{alg:1} was applied to phases $m=1,2,3,4$ sequentially.
The stage cost $\check{U}(x_{k},u_{k})$ is a quadratic form of state
$x_{k}$ and action $u_{k}$: 
\begin{equation}
\check{U}(x_{k},u_{k})=x_{k}^{T}R_{x}x_{k}+u_{k}^{T}R_{u}u_{k},\label{eq:utility-function}
\end{equation}
where $R_{x}\in\mathbb{R}^{2}$ and $R_{u}\in\mathbb{R}^{3}$ were
positive definite matrices. Specifically, $R_{x}=diag(1,1)$ and $R_{u}=diag(0.1,0.2,0.1)$
were used in our implementation. In the experimental results we chose
a quadratic stage cost \eqref{eq:utility-function}, which meets the
requirement of Assumption 1. But note however, our results in previous
sections apply to more general forms of stage cost functions. The
minimum memory buffer size $N_{b}$ was 20. During training, a small
Gaussian noise ($1\%$ of the initial impedance) was added to the
action output $u_{k}=h^{(i)}(x_{k})$ to create samples to solve \eqref{eq:policy-evaluation}.
The basis functions are $\phi(x_{k},u_{k})=[x(1)_{k}^{2},x(1)_{k}x(2)_{k},x(1)_{k}u(1)_{k},x(1)_{k}u(2)_{k},x(1)_{k}u(3)_{k},\allowbreak$$x(2)_{k}^{2},x(2)_{k}u(1)_{k},x(2)_{k}u(2)_{k},x(2)_{k}u(3)_{k},u(1)_{k}^{2},u(2)_{k}^{2},\allowbreak u(3)_{k}^{2},x(1)_{k}^{2}x(2)_{k},x(1)_{k}^{2}u(1)_{k},x(1)_{k}^{2}u(2)_{k}]^{T},$
where $x(1)_{k}$ denotes the first element of $x_{k}$, and so on. 

We define an experimental trial as follows. A trial started from gait
cycle $k=0$ until a success or failure status was reached. At the
beginning of each trial, the FS-IC was assigned with random initial
IC parameter $I_{0}$ as in \eqref{eq:imp}. The adaptive optimal
control objective for FPI is to make state $x_{k}$ approach zero,
i.e., the peak error $\text{\ensuremath{\Delta}}P_{k}$ and duration
error $\text{\ensuremath{\Delta}}D_{k}$ for all four phases approach
zero. We define upper bounds $P^{u}$ and $D^{u}$ and lower bounds
$P^{l}$ and $D^{l}$, and their values are identical to those in
\cite[Table I]{Wen2017}. Specifically, upper bounds $P^{u}$ and
$D^{u}$ are safety bounds for the robotic knee, i.e., $|\Delta P_{k}|\leq P^{u}$
and $|\Delta D_{k}|\leq D^{u}$ must hold during tuning. Lower bounds
$P^{l}$ and $D^{l}$ were used to determine whether a trial was successful:
the current trial is successful if $|\Delta P_{k}|<P^{l}$ and $|\Delta D_{k}|<D^{l}$
hold for 10 consecutive gait cycles before reaching the limit of 500
gait cycles; otherwise it is failed. The maximum memory buffer size
$N$ in Algorithm \ref{alg:1} was 100. The results in Subsections
\ref{subsec:ABPI-Learning-Performance} and \ref{subsec:Comparison-Study}
are based on 30 simulation trials. The success rate was the percentage
of successful trials out of 30 trials.

We used two performance metrics in the experiments: the learning success
rate as defined in Subsection \ref{subsec:Algorithm-and-Experiment},
and tuning time measured by the number of gait cycles (samples) needed
for a trial to meet success criteria. Tuning time also reflects on
data efficiency.

In summary, we conducted 30 testing trials following the procedure
below for each configuration of FPI to evaluate its learning performance
as reported in Tables \ref{tab:BPI-Tuner-Specification-onPolicy},
\ref{tab:Comparison} and \ref{tab:BPI-Tuner-Specification}. 

\begin{table}[b]
\caption{FPI Tuner Performance under Batch Mode\label{tab:BPI-Tuner-Specification-onPolicy}}

\begin{centering}
\begin{tabular}{>{\raggedright}m{0.2\columnwidth}>{\raggedright}m{0.2\columnwidth}l>{\raggedright}m{0.2\columnwidth}}
\toprule 
$N_{b}$ & Options{*} & Success Rate & Tuning Time

(gait cycles)\tabularnewline
\midrule
20 (Fixed) & \multirow{3}{0.2\columnwidth}{(A)(A)(A)(A)} & 76\% (23/30) & 93.4\textpm 13.6\tabularnewline
40 (Fixed) &  & 87\% (26/30) & 170.5\textpm 22.8\tabularnewline
100 (Fixed) &  & 100\% (30/30) & 428.6\textpm 52.2\tabularnewline
\midrule
20-40 (Ad.) & \multirow{2}{0.2\columnwidth}{(B)(A)(A)(A)} & 93\% (28/30) & 107.6\textpm 12.4\tabularnewline
40-100 (Ad.) &  & 100\% (30/30) & 268.0\textpm 22.5\tabularnewline
\bottomrule
\end{tabular}
\par\end{centering}
{*}refer to Table \ref{tab:Available-Options}. Ad.: adaptive.
\end{table}

\begin{table*}[t]
\caption{Performance Comparisons of prosthesis control\label{tab:Comparison} }

\begin{centering}
\begin{tabular}{lcccccccc}
\toprule 
\multicolumn{1}{c}{} & \multicolumn{2}{c}{FPI} & \multicolumn{2}{c}{GPI \cite{Liu2015c}} & \multicolumn{2}{c}{NFQCA \cite{Hafner2011}} & \multicolumn{2}{c}{dHDP \cite{Wen2017}}\tabularnewline
\multicolumn{1}{c}{} & SR & TT & SR & TT & SR & TT & SR & TT\tabularnewline
\midrule
Noise free & 93\% (28/30) & 107\textpm 12 & 53\% (16/30) & 384\textpm 33 & 47\% (14/30) & 213\textpm 48 & 73\% (22/30) & 323\textpm 136\tabularnewline
Uniform 5\% actuator & 90\% (27/30) & 106\textpm 17 & 53\% (16/30) & 402\textpm 37 & 47\% (14/30) & 218\textpm 48 & 70\% (21/30) & 332\textpm 124\tabularnewline
Uniform 10\% actuator & 83\% (25/30) & 112\textpm 19 & 53\% (16/30) & 401\textpm 36 & 40\% (12/30) & 226\textpm 54 & 73\% (22/30) & 348\textpm 141\tabularnewline
Uniform 5\% sensor & 83\% (25/30) & 105\textpm 15 & 50\% (15/30) & 384\textpm 33 & 43\% (13/30) & 220\textpm 50 & 73\% (22/30) & 326\textpm 122\tabularnewline
Uniform 10\% sensor & 80\% (24/30) & 128\textpm 21 & 43\% (13/30) & 421\textpm 28 & 33\% (10/30) & 223\textpm 51 & 70\% (21/30) & 342\textpm 138\tabularnewline
TF Human Subject 1 & 77\% (23/30) & 147\textpm 22 & 43\% (13/30) & 459\textpm 32 & 30\% (9/30) & 225\textpm 47 & 70\% (21/30) & 350\textpm 126\tabularnewline
TF Human Subject 2 & 80\% (24/30) & 142\textpm 17 & 40\% (12/30) & 456\textpm 41 & 36\% (11/30) & 245\textpm 53 & 70\% (21/30) & 361\textpm 129\tabularnewline
\bottomrule
\end{tabular}
\par\end{centering}
\medskip{}

FPI: proposed flexible policy iteration; GPI: generalized policy iteration;
NFQCA: neural fitted Q with continuous actions; dHDP: direct heuristic
dynamic programming; SR: Success rate for 30 trials; TT: Tuning Time,
which is the number of gait cycles to success.
\end{table*}

\subsection{FPI Batch Mode Evaluation\label{subsec:ABPI-Learning-Performance}}

We first evaluated the performance of FPI under its simplest form,
the batch mode where the entire batch ($N_{b}$ samples) was generated
under the policy to be evaluated (Setting 2(A) in Table \ref{tab:Available-Options}),
and neither PER nor supplemental value was considered. Table \ref{tab:BPI-Tuner-Specification-onPolicy}
summarizes the performance of FPI in batch mode with different batch
sizes. In our experiments we observed that the both the success rate
and tuning time rose as more samples (i.e. larger batch size $N_{b}$)
are used for policy evaluation. Table \ref{tab:BPI-Tuner-Specification-onPolicy}
also shows that, under Setting 2(A), adaptive batch mode improves
both success rates and tuning time over fixed batch mode.

\subsection{Comparisons with Other Methods\label{subsec:Comparison-Study}}

We now conduct a comparison study between FPI and three other popular
RL algorithms. These RL algorithms include generalized policy iteration
(GPI) \cite{Liu2015c}, neural fitted Q with continuous action (NFQCA)
\cite{Hafner2011} and our previous direct heuristic dynamic programming
(dHDP) implementation \cite{Wen2017}. GPI is an iterative RL algorithm
that contains policy iteration and value iteration as special cases.
To be specific, when the max value update index $N_{i}=0$, it reduces
to value iteration; when $N_{i}\rightarrow\infty$, it becomes policy
iteration. NFQCA and dHDP are two configurations similar in the sense
that both have features resemble SARSA and TD learning. According
to \cite{Hafner2011}, NFQCA can be seen as the batch version of dHDP. 

\begin{figure}[t]
\includegraphics[width=0.9\columnwidth]{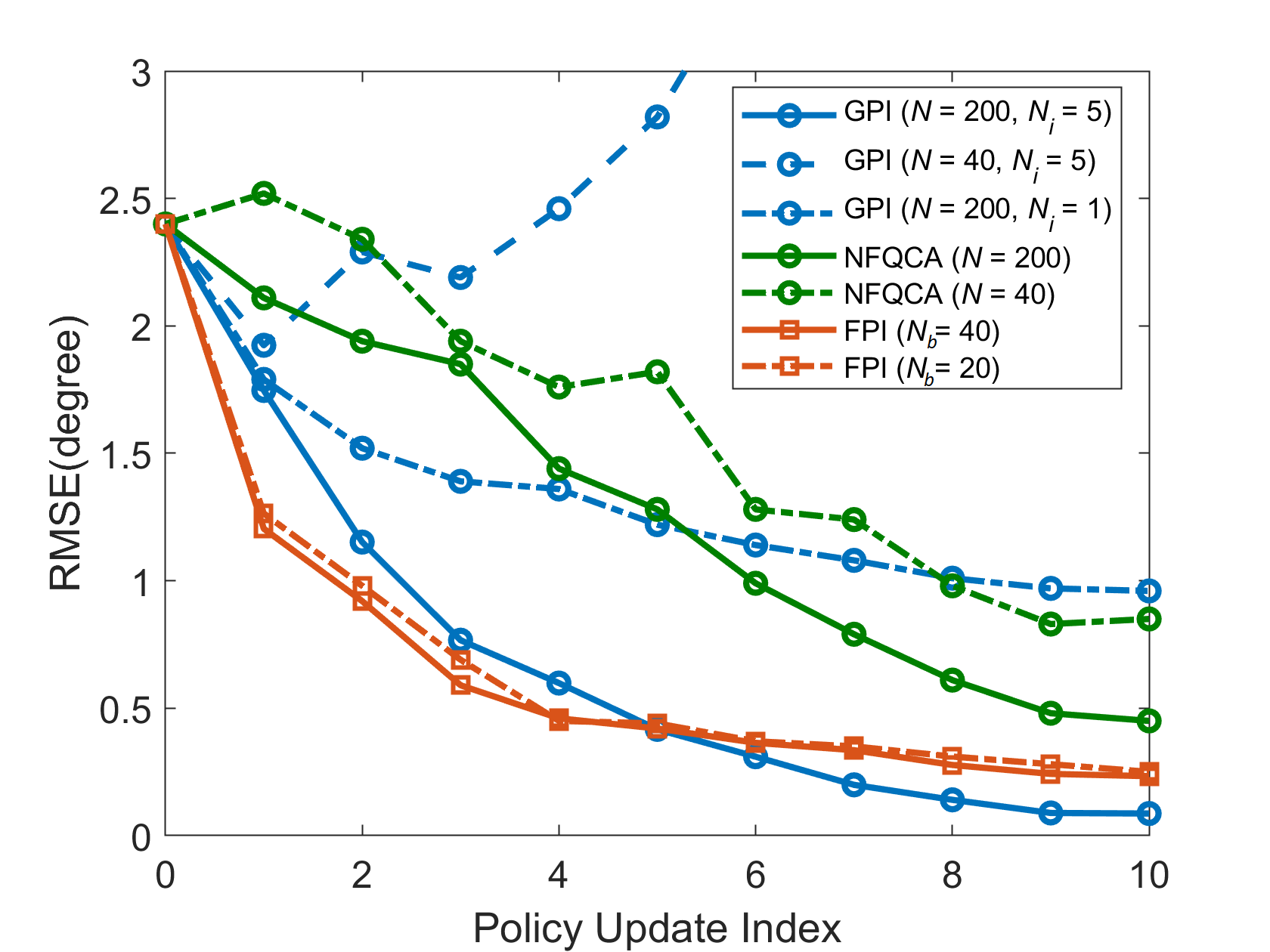}

\caption{Comparison of the RMSEs between controlled knee profiles and target
profiles using FPI, GPI and NFQCA under the same stage cost \eqref{eq:utility-function}.
\label{fig:Comparison-BPI-GPI-NFQCA}}
\end{figure}

To make a fair comparison between FPI and the other three RL algorithms,
we made FPI run under batch mode with neither PER nor supplemental
value involved. Specifically, results in Table \ref{tab:Comparison}
were based on an adaptive batch size $N_{b}$ between 20 and 40 (i.e.,
Settings (B)(A)(A)(A) in Table \ref{tab:Available-Options}), and
results in Fig. \ref{fig:Comparison-BPI-GPI-NFQCA} used a fixed $N_{b}$
of either 20 or 40 (i.e., Settings (A)(A)(A)(A) in Table \ref{tab:Available-Options}).

Before the comparison study, we first validated our implementations
of GPI, NFQCA and dHDP using examples from \cite{Wen2017,Hafner2011,Liu2015c},
respectively. We were able to reproduce the reported results in the
respective papers. For GPI, $N$ and $N_{i}$ were set equal to $p$
and $N_{i}$ as described in \cite{Liu2015c}, respectively. GPI's
critic network (CNN) and the action network (ANN) were chosen as three-layer
back-propagation networks with the structures of 2\textendash 8\textendash 1
and 2-8-3, respectively. For NFQCA, $N$ was equivalent to the pattern
set size $\#D$ in \cite{Hafner2011}. For both NFQCA and dHDP, CNN
and ANN were chosen as 5-8-1 and 2-8-3 respectively. Notice that the
number of neurons at the input layers are different, because NFQCA
and dHDP approximate the state action value function $Q(x_{k},u_{k})$
while GPI approximates $V(x_{k})$. To summarize, an effort was made
to make the comparisons fair. For example, FPI's batch sample size
$N_{b}$ was equivalent to GPI's and NFQCA's $N$, thus the maximum
$N_{b}$ (FPI), $N$ (GPI) and $N$ (NFQCA) were all set to 40 gait
cycles in Table \ref{tab:Comparison}.

Table \ref{tab:Comparison} shows a systematic comparison of the four
algorithms under various noise conditions. Artificially generated
noise and noise based on variations of human subject movement profiles
were used in the comparisons. To be specific, sensor noise and actuator
noise are uniform noise that are added to the states $x_{k}$ and
actions $u_{k}$, respectively. In the last two rows, human variances
collected from two amputee subjects TF1 and TF2 were introduced to
the simulations, which would affect the states $x_{k}$. Under all
noise conditions, FPI outperformed the other three existing algorithms
in terms of both success rate and tuning time. 

Fig. \ref{fig:Comparison-BPI-GPI-NFQCA} compares the root-mean-square
errors (RMSEs) between target knee angle profile and actual knee angle
profile using FPI, GPI and NFQCA. Note that when we used a suggested
parameter setting of ($N=40,N_{i}=5$) in GPI \cite{Liu2015c}, the
RMSE increased after a few iterations. Also note from Fig. \ref{fig:Comparison-BPI-GPI-NFQCA}
that, GPI may achieve a similar performance as the FPI but it required
a much larger sample size of $N=200$ than FPI's. 

\subsection{FPI Incremental Mode Evaluation}

We now evaluate FPI under incremental mode to further study FPI's
data and time efficiency. Both PER and learning from supplemental
value, two of the innovative features of FPI, can be employed in this
mode.

To obtain supplemental value $\mathcal{V}$ in \eqref{eq:augmented Q}
for the last row result in Table \ref{tab:BPI-Tuner-Specification},
we trained an FPI agent for just one trial in OpenSim under the same
settings as those in the first row of Table \eqref{tab:BPI-Tuner-Specification-onPolicy}
(Settings (A)(A)(A)(A) in Table \ref{tab:Available-Options} and $N_{b}=20$).
Then supplemental value $\mathcal{V}$ is obtained from $(x_{k})=\underset{u_{k}}{\min}Q_{f}(x_{k},u_{k})$
where $Q_{f}(x_{k},u_{k})$ the final approximate value function after
Algorithm 1 is terminated. 

\begin{table}[h]
\caption{FPI Tuner Performance under Incremental Mode\label{tab:BPI-Tuner-Specification}}

\begin{centering}
\begin{tabular}{>{\raggedright}m{0.16\columnwidth}>{\raggedright}m{0.28\columnwidth}l>{\raggedright}m{0.2\columnwidth}}
\toprule 
Configuration & Options{*} & Success Rate & Tuning Time

(gait cycles)\tabularnewline
\midrule
ER & (A)(B)(A)(A) & 83\% (25/30) & 134.4\textpm 21.6\tabularnewline
PER & (A)(B)(B)(A) & 83\% (25/30) & 127.6\textpm 25.8\tabularnewline
PER+Supp value & (A)(B)(B)(B) & 90\% (27/30) & 103.3\textpm 15.1\tabularnewline
\bottomrule
\end{tabular}
\par\end{centering}
\medskip{}

{*}refer to Table \ref{tab:Available-Options}. ER: Experience Replay;
PER: Prioritized Experience Replay.
\end{table}

Table \ref{tab:BPI-Tuner-Specification} summarizes the performance
of FPI in incremental mode under three different configurations. ER
or PER reutilized past samples from the current trial for policy iteration
(Settings 2(B) in Table \ref{tab:Available-Options}). The first configuration
is the ER case without sample prioritization, i.e., $\lambda_{k}^{(i)}=1$
for all $k$. The second configurations prioritized the samples before
performing the policy evaluation. In both the first and the second
configurations (the first two rows in Table \ref{tab:BPI-Tuner-Specification}),
no supplemental value was used, i.e., $\mathcal{V}(x_{k})=0$ for
all $x_{k}.$ The third configuration (the third row in Table \ref{tab:BPI-Tuner-Specification})
utilized both prioritized samples and supplemental value. The supplemental
value $\mathcal{V}(x_{k})$ was obtained from training FPI with a
previous trial. In Table \ref{tab:BPI-Tuner-Specification}, the success
rate increases from 83\% to 90\% as the algorithm gets more complex
with PER and supplemental value. The results also suggest that the
introduction of sample prioritization and supplemental value improves
the data efficiency. Note that if the maximum number of gait cycles
was extended from 500 to 1000, then the success rate of all simulation
results in Table \ref{tab:BPI-Tuner-Specification} will be 100\%.

\begin{figure}[t]
\includegraphics[width=0.9\columnwidth]{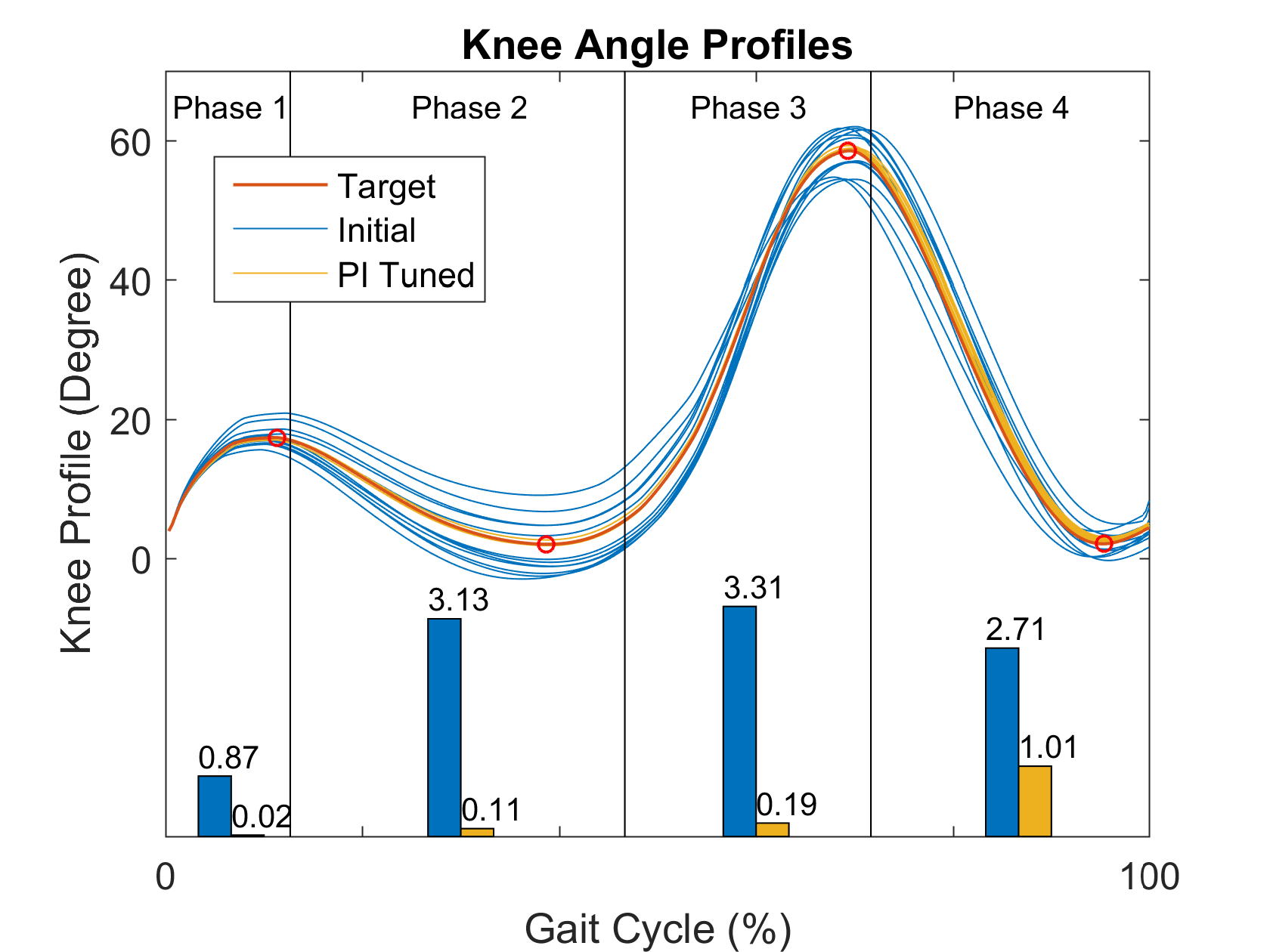}

\caption{Before-and-after FPI tuning of knee profiles of 15 randomly selected
trials. Top half: FPI enabled initial knee profiles (blue) approach
the target knee profile (red), with FPI enabled, final knee profiles
shown in yellow. Bottom half: The blue bars are the RMSEs between
the initial knee angle profiles and the target knee profile, and the
yellow bars are the RMSEs between the FPI tuned knee profiles and
the target profile. \label{fig:knee-profile}}
\end{figure}

A statistical summary of a 30 randomly initialized trials based on
the condition in row 1 of Table \ref{tab:BPI-Tuner-Specification}
is provided in Fig. \ref{fig:knee-profile} (bottom half panel). As
shown, after tuning, the proposed FPI algorithm successfully reduced
gait peak and duration errors.

\section{Conclusion\label{sec:Conclusion}}

We have proposed a new flexible policy iteration (FPI) algorithm aimed
at providing data and time efficient, high-dimensional control inputs
to configure a robotic knee with human in the loop. The FPI incorporates
previous samples and supplemental values during learning using prioritized
experience replay and an augmented policy evaluation. Our results
not only show qualitative properties of FPI as a stabilizing controller
and that it approaches approximate (sub)optimal solution, but also
include extensive simulation evaluations of control performance of
FPI under different implementation conditions. We also compared FPI
with other comparable algorithms, such as dHDP, NFQCA and GPI, which
further demonstrates the efficacy of FPI as a data and time efficient
learning controller. The FPI under batch mode became more efficient
when utilizing (prioritized) experience replay and previous knowledge.
Even though our application does not render itself as a big data problem,
but our results show that FPI has the capability of efficiently working
with a tight data budget. Specifically, FPI is capable of successfully
tuning the control parameters within 100\textquoteright s gait cycles
under various conditions (Tables II and III), which is an equivalent
of only a few minutes of walking time. Our results reported here represent
the state of the art in automatic configuration of powered prosthetic
knee devices. This result can potentially lead to practical use of
the FPI in clinics. In turn, this can significantly reduce health
care cost and improve the quality of life for the transfemoral amputee
population in the world. 

\bibliographystyle{IEEEtran}
\bibliography{main}

\end{document}